\documentclass[twocolumn,epjc3]{svjour3}  
\smartqed  % flush right qed marks, e.g. at end of proof
\RequirePackage{graphicx}
\journalname{Eur. Phys. J. A}
\usepackage{color}
\usepackage{amssymb}
\usepackage[dvipsnames]{xcolor}

\usepackage{tikz}
\usetikzlibrary{quantikz}

\usepackage{youngtab}
\usepackage{young}
\usepackage{ytableau}
\usepackage{bm}
\usepackage{ulem}
\usepackage[colorlinks=true,allcolors=Blue,pdfusetitle]{hyperref}

\usepackage{tabularx}
\usepackage{slashbox}

\usepackage[utf8]{inputenc}
\usepackage{fourier} 
\usepackage{array}
\usepackage{makecell}

\begin{document}

\title{Symmetry breaking/symmetry preserving circuits and symmetry restoration on quantum computers }
\subtitle{A quantum many-body perspective}


\author{Denis Lacroix\thanksref{e1,addr1}
        \and
        Edgar Andres Ruiz Guzman\thanksref{e2,addr1} 
        \and 
        Pooja Siwach\thanksref{e3,addr2}  %etc.
}

\thankstext{e1}{e-mail: denis.lacroix@ijclab.in2p3.fr}
\thankstext{e2}{e-mail: ruiz-guzman@ijclab.in2p3.fr}
\thankstext{e3}{e-mail: psiwach@physics.wisc.edu}


\institute{Universit\'e Paris-Saclay, CNRS/IN2P3, IJCLab, 91405 Orsay, France \label{addr1}
           \and       
Department of Physics, University of Wisconsin--Madison, Madison, Wisconsin 53706, USA \label{addr2}
%           \emph{Present Address:} if needed\label{addr3}
}

\date{Received: date / Accepted: date}

\maketitle

\begin{abstract}
We discuss here some aspects related to the symmetries of a quantum many-body problem when trying to treat it on a quantum computer. Several features related to symmetry conservation, symmetry breaking, and possible symmetry restoration are reviewed. After briefly discussing some of the standard symmetries relevant for many-particle systems, we discuss the advantage of encoding some symmetries directly in quantum ansätze, especially to reduce the quantum register size. It is, however, well-known that the use of symmetry-breaking states can also be a unique way to incorporate specific internal correlations when a spontaneous symmetry breaking occurs. These aspects are discussed in the quantum
computing context. Ultimately, an accurate description of quantum systems can be achieved only when the initially broken symmetries are properly restored.
We review several methods explored previously to perform symmetry restoration on a quantum computer, for instance, the ones based on symmetry filtering by quantum phase estimation and by an iterative independent set of Hadamard tests. We propose novel methods that pave the new directions to perform symmetry restoration, like those based on the purification of the state employing the linear combination of unitaries (LCU) approach.
\end{abstract}
%
% Computer program descriptions must contain the following
% PROGRAM SUMMARY AND SPECIFICATIONS.
\noindent
\section{Introduction}
\label{intro}

Recognizing symmetries is an important aspect when trying to solve a quantum many-body problem and, more generally, a physical one. Symmetries or approximate 
symmetries can generally be uncovered from the observation of a system. Symmetries usually lead to invariance, conservation laws, and regularities
\cite{Gro96}. When considering a quantum or classical complex problem, symmetries are usually of paramount importance to facilitate its resolution \cite{Wei95a,Wei95b}. Here, we focus on the case of quantum problems. 
The use of symmetries helps to reduce the complexity of the description
of a problem by focusing on the relevant part of the Hilbert space where the system is described \cite{Mes62,Kap75,Ham12}.  

We discuss the concept of symmetries in the context of quantum computing \cite{Fan19,Cao19,McA20,Bau20,Bha22} with a specific focus on the problem of strongly 
interacting (Fermi) systems. Different aspects of the use of symmetries are treated, two of them being the reduction in resources (number of gates, number of qubits) thanks to the consideration of symmetries in the fermion-to-qubit mapping and the interest in using symmetry-preserving ans\"atze. Although it is not the main focus of the present work, for completeness, we briefly discuss this mapping in \ref{sec:fermion}.

In some highly non-perturbative problems, it is sometimes useful to use symmetry-breaking states to describe complex correlations 
between particles \cite{Rin80,Bla86}. 
This technique, standardly employed to treat many-body systems in physics or quantum chemistry, has been much less explored in quantum computing. One prerequisite to accurately describing strongly correlated systems is to restore the symmetry once it has been broken. 
A typical illustration of systems where symmetry-breaking plays 
a key role is the atomic nuclei. In this system, it might be advantageous to break particle number, parity, time-reversal, or rotational symmetry to grasp specific internal correlations \cite{Ben03,Rob18,She19}. Most applications today are performed in the Energy Density Functional (EDF) framework where Symmetry-Breaking (SB) followed by Symmetry-Restoration (SR) has become a standard tool to describe the structure of nuclei, including the heaviest ones. More recently, this method has been exported to the so-called ``ab-initio" theories as a potential candidate to treat open-shell systems \cite{Lac12,Gam12,Dug17,Qiu17,Qiu19,Rip17,Rip18,Fro21a,Fro21b,Fro21c}. Using an SR-SB strategy in various ab-initio methods where one starts from a bare Hamiltonian with two-body and eventually three-body interactions is numerically very demanding on a classical computer and still restricts its applications. Such studies are focused on nuclear properties. Some attempts also exist to use the same strategy for dynamic \cite{Reg19}, or statistical problems \cite{Ese93,Gam13}. Due to the numerical complexity, only schematic problems can currently be solved in these cases. Quantum computers might potentially open new perspectives for applications of ab-initio methods to a broader class of systems and for taking advantage of symmetry breaking / symmetry restoration in dynamical or statistical problems. 

Besides the purely ``many-body problem" motivation, symmetry restoration can be a valuable instrument in the quantum computing toolkit. The restoration might, for instance, be crucial to control the noise leading to accidentally broken symmetries \cite{McA19}, or might be of particular interest in the quantum machine learning context to treat problems that are invariant under 
permutation (see for example \cite{Sch22}).

The present article aims to give a concise overview of the use of symmetries in quantum computers. We first discuss some operators associated with the most standard symmetries and their encodings on qubits. Different aspects, from enforcing certain symmetries, breaking these symmetries, and restoring symmetries, are explored as well.

Special attention has been put herey on restoring broken symmetries that are important for many-body systems and can also be helpful for broader applications like reducing the errors on quantum computers or in unstructured search. We discuss the notion of projection onto a subspace of the Hilbert space where symmetries are respected and the oracles associated with the projection operators.  
Several methods to perform symmetry restoration are presented; some of them, as far as we know, being new.

After a summary of the notations used here, we discuss in \sectionautorefname~\ref{sec:notation}
how some of the most common symmetry operators (particle number, total spin, or parity) are treated on a quantum computer. We then discuss in section \ref{sec:ansatz} the construction of wave functions that do or do not respect the symmetries. We underline in this section the advantage of not breaking symmetries in terms of quantum resources and physical problems. This section also illustrates how some standard states, like Coupled-Cluster or BCS states, can be constructed in quantum computers. 
In section \ref{sec:oracle}, we discuss the connections between projectors and quantum oracles that are generally used in the quantum search algorithm. Such a link will be helpful in one of the new approaches we propose here. 
We show in following sections (section \ref{sec:bs} to \ref{sec:amplitude})
that we have at our disposal a large variety of methods to perform symmetry restoration on quantum computers, some giving access to expectation values of observables, others giving the projected state directly. Some methods we discuss have been proposed previously, like those based on indirect measurements (section \ref{sec:indirect}). In contrast, others are new methods like the amplitude amplification methods (\ref{sec:amplitude}) or the direct implementation of projectors (section \ref{sec:lcu}).  

\section{Notations, conventions and simple discussions on symmetries}\label{sec:notation}

Let us consider a set of $N$ qubits associated to a basis 
denoted by $| s_{N-1}  \cdots s_{0} \rangle$ with $s_i =0$, $1$. We will sometime use the notations $| 0_j\rangle$ and $| 1_j \rangle$ for the states corresponding to the qubit $q_j$ for $j=0,\cdots,N-1$. These states, called 
natural basis (NB) hereafter, form a complete basis of size $2^N$. Any state of the NB can be written equivalently as $\bigotimes_j | s_j \rangle  = | n \rangle$ provided that $n=\sum_j s_{j} 2^{j}$, such that $[s_{N-1}  \cdots s_{0}]$ is the binary representation of the integer $n$ where  $0 \le n < 2^N$.  Considering a wave-function $| \Psi \rangle$, its most general decomposition in the NB is given by:
\begin{eqnarray}
| \Psi \rangle = \sum_{s_j \in\{0,1\}} \Psi_{s_{N-1}, \cdots, s_0 } | s_{N-1}  \cdots s_{0}  \rangle. \label{eq:wavegeneral}
\end{eqnarray}

\subsection{Some symmetries and their consequences}
\label{sec:symcons}

If a physical system respects some symmetries, the coefficients in Eq. (\ref{eq:wavegeneral}) follow specific properties associated with the symmetry.
We show below for some standard examples of symmetries in physics what is the consequence of respecting these symmetries on the wave-function. To make contact with physical problems, we consider particles with spins $1/2$. This problem can be mapped onto a quantum computer assuming that each particle is described by one qubit and that the states $| 0 \rangle_j$ and $| 1 \rangle_j$
correspond to spin up and spin down of a particle $j$, respectively. We introduce the three Pauli matrices 
\begin{eqnarray}
X_j = \left[ \begin{array}{rr}0&1\\ 1&0\end{array}\right]_j, ~
Y_j = i \left[\begin{array}{rr}0&- 1\\ 1&0\end{array}\right]_j,  ~
Z_j = \left[\begin{array}{rr}1&0\\ 0&-1\end{array}\right]_j  \label{eq:Pauli}
\end{eqnarray}
complemented by the identity denoted by $I_j$. The spin components 
of each particle $j$ are then given by ${\bf S}_j = \frac{1}{2} (X_j, Y_j, Z_j) $ while, 
for the total set of particles, we define the total spin operator ${\bf S} = \sum_j {\bf S}_j $.

In many physical problems, the hamiltonian $H$ commutes with both ${\bf S}^2$ and ${S_z}$, and eigenstates 
verify $S_z| \Psi \rangle = M | \Psi \rangle$ and ${\bf S}^2| \Psi \rangle = S(S+1) | \Psi \rangle$ (we take the convention below $\hbar=1$) where $M$ 
and $S$ are both integers (see below).  This implies strong constraints on the decomposition of $| \Psi \rangle$ given by Eq. (\ref{eq:wavegeneral}). To illustrate this aspect, we consider the two symmetries separately:
\begin{itemize}
 \item 
  {\bf Total spin azimuthal projection:}  we first consider eigenstates of $S_z = ( \sum_j Z_j)/2$. It is straightforward to show that 
  any state $| k \rangle$ of the NB is an eigenstate of $S_z$ with eigenvalue $M(k) = [n_0(k) - n_1(k)]/2$ where $n_0 (k)$ (resp. $n_1(k)$)
  is the number of $0$ (resp. $1$) in the binary representation of $k$. Since the total number of qubits is fixed, we also have the constraint $n_0 (k) + n_1 (k) = N$ leading to:
  \begin{eqnarray}
M(k) &=& \frac{N}{2} - n_1(k) =  n_0 (k) - \frac{N}{2}. \nonumber
\end{eqnarray}   
If we consider the subspace associated to a given value of $M$, this subspace is highly degenerated and the degeneracy corresponds to 
the number of states in the NB verifying $n_0 = N/2 - M$, i.e. $C^{n_0}_N$. A corollary of this is that the relevant Hilbert space for the problem with symmetry imposed is much smaller than $2^N$ and any state given by Eq. (\ref{eq:wavegeneral}) has at maximum $C^{n_0}_N$  non-zero components. In classical computers, one often takes advantage of such symmetries by considering only the subspace of states with the proper symmetry.

\begin{figure}
\begin{center}
 \includegraphics[width=\columnwidth] {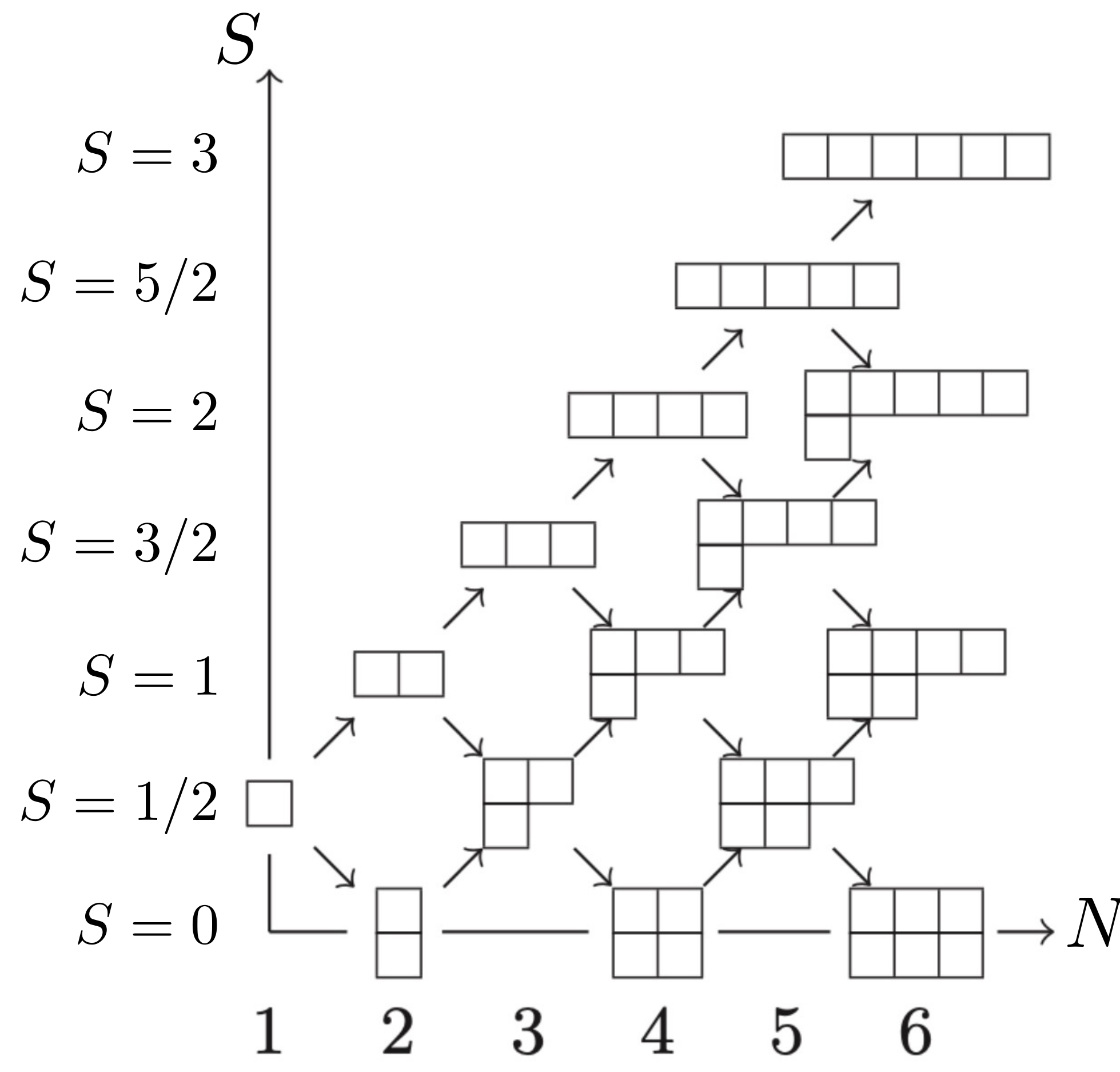} %0.8\linewidth
 \end{center}
\caption{Illustration of the different Young tableaux construction for different numbers of qubits $N$ (adapted from \cite{Hav19})). The Young tableaux for spin systems are restricted to those with only two horizontal rows with length $l_u$ (upper line) and $l_d$ (lower line)
and with the property $S = l_u - l_d \ge 0$ where $S$ is the total spin. Each line is composed of blocks (squares). The schematic illustration gives an iterative procedure to construct eigenstates of ${\bf S}^2$ by adding one particle (one qubit) at a time. In practice, symmetry preserving states can be constructed by assigning a qubit to each block with possible values $0$ and $1$. Then the eigenstate is symmetric with respect to the exchange of indices in the same horizontal column and antisymmetric for indices in blocks of the same vertical line. In the latter case, two $0$s or two $1$s are therefore impossible in the same vertical line.  
The state with only one horizontal line is therefore fully symmetric with respect to the exchange of indices. Illustration of few states are given in the text.}
\label{fig:spin}       
\end{figure}

  \item {\bf Total spin:} We now consider the effect of being an eigenstate of ${\bf S}^2$ given by the expression:
  \begin{eqnarray}
{\bf S}^2 &=& \frac{1}{4} \sum_{jl} \left( X_j X_l + Y_j Y_l + Z_j Z_l   \right). \label{eq:s2}
\end{eqnarray}
This operator is directly connected to the group of permutation \cite{Mes62,Kap75,Ham12}. The link becomes explicit from 
the alternative expression of ${\bf S}^2$ \cite{Dir35,Low70}:
\begin{eqnarray}
{\bf S}^2 &=&  \frac{N (4-N)}{4} I +  \sum_{j<l, l=0}^{N-1}  P_{jl}, \label{eq:transpos}
\end{eqnarray}
where $P_{jl}$ is the transposition operator given by
$P_{jl} = \frac{1}{2} (I+X_jX_{l} +Y_jY_{l} +Z_jZ_{l})$. This operator permutes two qubits, i.e. $P_{jl}| s_j s_{l}\rangle = | s_{l} s_j \rangle$.
Because of Eq. (\ref{eq:transpos}), eigenstates of ${\bf S}^2$ have specific symmetries with respect to the permutation of qubit indices that can be for instance illustrated using the Young tableaux shown in Fig. \ref{fig:spin}. As an illustration of the use of Young tableau, we can 
construct  the eigenstates of $({\bf S}^2, S_z)$ denoted by $|S,M \rangle$ with $S \le N/2$ and $M=-S, \cdots, +S$. 
For $N=2$, remembering that $M$ is related to the number of $0$ (or $1$), we have:
  \begin{eqnarray}
  \begin{array}{lclllll}
\Yvcentermath1
&   {\scriptsize\young(00)} &  \rightarrow   &   | 00 \rangle  & (M=1) \\
 {\small (S= 1)}    &
 {\scriptsize\young(01)}    &\rightarrow   &  \frac{1}{\sqrt{2}} \left(  | 10 \rangle + |01 \rangle \right)   & (M=0)  \\
&    {\scriptsize\young(11)} & \rightarrow    &   | 11 \rangle   &(M=-1) \\
\Yvcentermath1 
 {\small (S= 0)}  &  \Yvcentermath1 {\scriptsize\young(0,1)} &   \rightarrow  & \frac{1}{\sqrt{2}} \left(  | 10 \rangle - |01 \rangle \right) & (M = 0)\nonumber
 \end{array} 
 \end{eqnarray}        
Eigenstates of $N=3$ can then be constructed iteratively starting from the $N=2$ case 
or directly by considering all possible allowed combinations of $0$ and $1$ in different qubits.  
The iterative procedure can be continued up to a given number $N$ of qubits and an eigenstate will correspond to a path in the figure shown in Fig. \ref{fig:spin}. 
It could be shown that all eigenstates of $({\bf S}^2, S_z)$ can be constructed in this way. 
Such states also form a complete basis of the Hilbert space that we call total spin basis (TSB). It is worth mentioning that the subspace corresponding
to the $(S,M)$ eigenvalues is degenerate. The degeneracy is linked to the number of paths leading to a given Young tableau shown in Fig. \ref{fig:spin}. However, each eigenstate in a given block is linked to a single path. This property was recently used in Ref.~\cite{Siw21} to construct specific eigenstates on a quantum computer.       

%To give few example in $N=3$ 
%\begin{eqnarray}  
%\begin{array}{ll}
% \scriptsize\young(011)       &  \rightarrow  | 3/2, -1/2 \rangle = \frac{1}{\sqrt  3} \left( | 110 \rangle + | 101 \rangle + |110 \rangle \right)   \\
% {\scriptsize\young(00,1)}    &  \rightarrow   | 1/2, +1/2 \rangle =   \sqrt{ \frac{2}{3}} | 100  \rangle  - \frac{1}{\sqrt{6}} (| 010 \rangle + |001 \rangle )  \\ 
% \end{array}  \nonumber
%\end{eqnarray}  
%\begin{eqnarray}
 % \Yvcentermath1 {\scriptsize\young(12,3)} &\rightarrow&  
%\begin{array}{lll}
%{\scriptsize\young(++,-)}    &  \rightarrow   | 1/2, +1/2 \rangle_{[3]} &=   \sqrt{ \frac{2}{3}} | 100  \rangle  - \frac{1}{\sqrt{6}} (| 010 \rangle + |001 \rangle )  \\ 
%&&=  \\
%{\scriptsize\young(+-,-)}      &  \rightarrow  | 3/2, -1/2 \rangle_{[3]} &=\sqrt{ \frac{2}{3}} | 011  \rangle  - \frac{1}{\sqrt{6}} (| 110\rangle + |101 \rangle ) 
 % \\ 
%&&=
%\end{array} 
% \end{eqnarray}  

We finally mention that the TSB can be used as an alternative basis to the NB for quantum computing. This is actually the essence of the   
permutational quantum computing (PQC) introduced in Refs.~\cite{Mar02,Mar05} and further discussed in Ref.~\cite{Jor10}.  

\item {\bf Qubit Parity:} The qubit parity operator \cite{McA19}, denoted by $\pi$ has analogy with the parity of a classical string and/or parity of a system of spins. It has  two eigenvalues that dissociate the states of the NB into two subsets: those with odd and those with even number of $1$. A possible choice for such operator is simply to take $\pi = \bigotimes_{j} Z_j$. When acting on a state of the NB, we see that we have:
\begin{eqnarray}
\pi | s_{N-1}  \cdots s_{0}  \rangle &=& (-1)^{\sum_j {s_j}}| s_{N-1}  \cdots s_{0}  \rangle,
\end{eqnarray}
showing that these states are eigenstates of the operator $\pi$ with eigenvalues $+1$ or $-1$ if the number of $1$ is even or odd, respectively.  
If, for instance, the state (\ref{eq:wavegeneral}) is an eigenstate of $\pi$ with parity $+1$, this automatically implies that the components 
on the NB states with parity $-1$ are zero and we can simplify the problem by restricting the Hilbert space to the even parity block.     

%For a classical bit string, the parity is equal to $0$ (resp. $1$) if the number of $1$ is even (resp. odd). The Quantum parity operator  $\pi$ is 
%defined as 
%\begin{eqnarray}
%\pi & = & 
%\end{eqnarray}  
   
\end{itemize}

\section{Symmetry preserving vs symmetry breaking quantum ans\"atze}
\label{sec:ansatz}

In classical computers, symmetries are often used to reduce the complexity of a problem by reducing the size of the Hilbert space considered, i.e., by considering only those states as possible solutions that already have the proper symmetries. 

In quantum computers, the direct use of the JWT (see \ref{sec:fermion}), where a qubit is assigned to each single-particle state, leads to the utilization of the entire Hilbert space; this could be seen as a waste of resources since we know that the relevant states belong to a subspace that respects the symmetries. To tackle this waste, we have explored two different approaches. We can alter the considered ansatz so that the trial wave-function respects the symmetries. This usually reduces the number of parameters/quantum gates considered. Or, we can consider the symmetries directly in encoding the physical problem, and by doing so, we reduce the number of needed qubits.

\subsection{Reduction of the circuit complexity using symmetries: an illustration}

\begin{figure}[htbp]  
\centering
(a) \\
\begin{tikzpicture}
  \node[scale=0.8] { %scale=0.75
\begin{quantikz} 
 %\lstick{\ket{0}}  & 
 \push{} & \gate{A_1} &\targ{}  &\gate{R_Z(\theta_1)}  &\ctrl{1} &\push{}    &\targ{} & \gate{A_3}  & \push{} \qw \\
 %\lstick{ \ket{0}}  & 
 \push{} &\gate{A_2} &  \ctrl{-1}  &\gate{R_Z(\theta_2)}  & \targ{}  &\gate{R_Z(\theta_3)}    & \ctrl{-1} &  \gate{A_4} & \push{}  \qw
\end{quantikz}
}; 
\end{tikzpicture}\\
(b) \\
\begin{tikzpicture}
  \node[scale=0.8]{ %scale=0.75
\begin{quantikz} 
 %\lstick{ \ket{0}}  & 
 \push{} &  \ctrl{1}  &\gate{R^{\dagger}(\theta_1, \theta_2)}  & \targ{}  & \gate{R(\theta_1, \theta_2)}    & \ctrl{1} &   \push{}  \qw \\
 %\lstick{\ket{0}}  & 
 \push{}  &\targ{}  & \push{} &\ctrl{-1} &\push{}    &\targ{} &  \push{} \qw 
\end{quantikz}
};
\end{tikzpicture} \\
(c) \\
\begin{tikzpicture}
  \node[scale=0.8]{ %scale=0.75
\begin{quantikz} 
 %\lstick{ \ket{0}}  & 
 \push{} &  \ctrl{1} & \gate{X} &\gate{R^{\dagger}(\theta_1, \theta_2)}  & \targ{}  & \gate{R(\theta_1, \theta_2)}  & \gate{X}   & \ctrl{1} &   \push{}  \qw \\
 %\lstick{\ket{0}}  & 
 \push{}  &\targ{}  & \push{} & \push{} &\octrl{-1} &\push{} &\push{} &\targ{} &  \push{} \qw 
\end{quantikz}
};
\end{tikzpicture} \\
(d) \\
\begin{tikzpicture}
  \node[scale=0.8]{ %scale=0.75
\begin{quantikz} 
 %\lstick{ \ket{0}}  & 
 %\push{} &  \gate{X} &\gate{R^{\dagger}(\theta_1, \theta_2)}  & \gate{X}  & %\gate{R(\theta_1, \theta_2)}  & \gate{X}   &   \push{}  \qw \\
 %\lstick{\ket{0}}  & 
 %\push{}  &\targ{}  & \push{} & \push{} &\octrl{-1} &\push{} &\push{} %&\targ{} &  \push{} \qw 
\push{} & \gate{X} & \gate{R^{\dagger}(\theta_1, \theta_2)} & \gate{X} & \gate{R(\theta_1, \theta_2)} & \gate{X} & \push{} 
\end{quantikz}
};
\end{tikzpicture} \\

    \caption{ Panel (a): Circuit proposed in Ref. \cite{Vat04} to simulate a general unitary matrix acting on $2$ qubits and that minimizes the number of CNOTs. 
    Each $A_{i}$ operator can be decomposed as $A_i = R_{Z}(\alpha_i) R_Y(\gamma_i)  R_{Z}(\delta_i)$. We use here the standard conventions for rotations $R_O(\alpha) = e^{-i \alpha O/2}$. Each $A_i$ has 3 parameters and consequently, the circuit shown in panel (a) has $15$ parameters. In all the circuits, we follow the qiskit convention \cite{Abr19} where the uppermost qubit corresponds to the least significant bit in the binary notation.  
    Panel (b): illustration of circuit corresponding to the unitary transformation given by Eq. (\ref{eq:01only}) \cite{Gar20}. 
    In this circuit $R(\theta_1, \theta_2) = R_Z (\theta_2 + \pi) R_Y(\theta_1+\pi/2)$. Panel (c): circuit used to obtain the    
    unitary transformation in Eq. (\ref{eq:0011only}). Panel (d): circuit used to obtain the unitary transformation Eq. (\ref{eq:reduced}). 
    The circuits shown here and the ones in the following are made using
the quantikz package~\cite{Kay18}.}
    \label{fig:circsym2}
\end{figure}
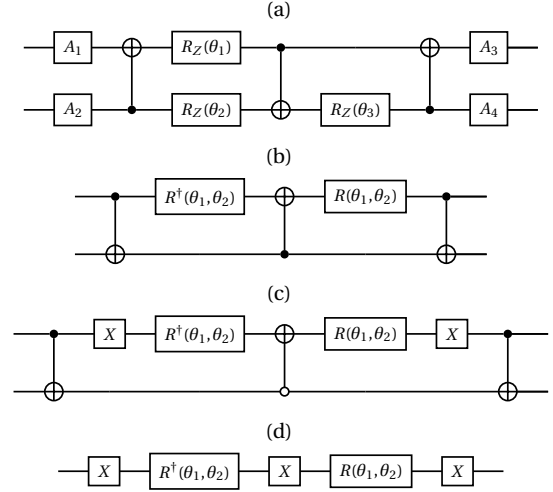

Let us consider a simple case of two qubits to illustrate the concept of symmetry preserving state. The Hilbert space is of dimension $2^2$ and contains the states $\{ | 00 \rangle, | 01 \rangle , | 10 \rangle, | 11 \rangle \}$. On a quantum computer, assuming that all qubits 
are set to $|0\rangle$ at the beginning of the calculation, a trial state vector is given by 
\begin{eqnarray}
| \Psi(\bm{ \theta})\rangle &=& U(\bm{ \theta}) | - \rangle , \label{eq:gen}
\end{eqnarray}     
where $U(\bm{ \theta})$ is a unitary operator while $\bm{ \theta} = \{ \theta_1, \cdots , \theta_\Omega\}$ stands  
for a set of parameters.  Because of the unitarity of $U$, a fully unrestricted version of this matrix can have at maximum $15$ independent real parameters.  A general circuit with minimal number of CNOT gates to generate $U$ was discussed in Ref. \cite{Vat04}. This circuit is shown in Fig. \ref{fig:circsym2}-a.

We now illustrate the effect of imposing symmetries on $U$. Let us first assume that we want a unitary transformation that preserves the Hamming weights \cite{HammingW} presented in the initial wave function (i.e. $S_z$ conserving case). In a many-body problem, this could be seen as the
particle number conservation.  In our two qubit problem this implies that $|01\rangle$ and $|10 \rangle$ can mix up through a SU(2) unitary transformation. The matrix $U$ can then be written into the following block diagonal form with only two parameters (see for instance the discussion in 
\cite{Gar20}):  
 \begin{eqnarray} 
 U(\theta_1, \theta_2)   &=&
 \left[
  \begin{array}{cccc}
1 & 0&0&0 \\
0 &  \cos(\theta_1)  & e^{i\theta_2} \sin(\theta_1)& 0\\
0 &  e^{-i\theta_2} \sin(\theta_1)& -\cos(\theta_1)  & 0 \\
0 & 0 & 0 & 1\\
% \cos( \Delta t \alpha )    &  0 & 0 & -i \sin(\sqrt{\Delta t \alpha})   \\
% 0     &  1 & 0 &  0   \\
% 0     &  0 & 1  &0  \\
% -i \sin(\sqrt{\Delta t \alpha})     & 0 & 0 &  \cos( \Delta t \alpha )
\end{array}
\right] . \label{eq:01only}
\end{eqnarray}
The number of parameters is therefore significantly reduced. In parallel, as illustrated in Fig. \ref{fig:circsym2}-b. The complexity of the circuit 
is also reduced. In the context of many-body systems, any state with one particle can be constructed from:
\begin{eqnarray}
| \Psi(\theta_1, \theta_2) \rangle &=& U(\theta_1, \theta_2) \left( I \otimes X \right) |-\rangle \nonumber \\
&=&  \cos(\theta_1) |01 \rangle + e^{-i\theta_2} \sin(\theta_1) | 10 \rangle. \label{eq:state01}
\end{eqnarray}

Another symmetry that could be imposed on the circuit is the parity. Assuming that only states with same parity can mix with each other, we then have 
two separated blocks, the even parity block $\{ | 00 \rangle ,  | 11 \rangle \}$  and the odd parity block $\{ | 01 \rangle ,  | 10 \rangle \}$. The first block is of special interest for treating superfluid systems. We focus here on this block and consider the specific case where $U$ leaves the odd parity states unchanged, i.e.:
 \begin{eqnarray} 
 U(\theta_1, \theta_2)  &=&
 \left[
  \begin{array}{cccc}
\cos(\theta_1)  & 0&0&e^{i\theta_2} \sin(\theta_1) \\
0 & 1 & 0 & 0\\
0 & 0 &1 & 0 \\
e^{-i\theta_2} \sin(\theta_1) & 0 & 0 & -\cos(\theta_1)   \\
% \cos( \Delta t \alpha )    &  0 & 0 & -i \sin(\sqrt{\Delta t \alpha})   \\
% 0     &  1 & 0 &  0   \\
% 0     &  0 & 1  &0  \\
% -i \sin(\sqrt{\Delta t \alpha})     & 0 & 0 &  \cos( \Delta t \alpha )
\end{array}
\right]. \label{eq:0011only}
\end{eqnarray}
The circuit that performs the above unitary transformation is shown in Fig. \ref{fig:circsym2}-c. We see in particular that the present circuit acting on the qubit vacuum will give:
\begin{eqnarray}
U(\theta_1, \theta_2)|-\rangle &=&  \cos(\theta_1) |00 \rangle + e^{-i\theta_2} \sin(\theta_1) | 11 \rangle. \label{eq:genbell}
\end{eqnarray} 
The resulting state can be seen as a generalized Bell state. As we will discuss further below for many-body systems, this state mixes the
components with $0$ and $2$ particle numbers in Fock space.   

The two examples above illustrate how symmetries' imposition can significantly reduce the circuit length. In the two illustrations aforementioned, the targeted states in Eqs. (\ref{eq:state01}) and (\ref{eq:genbell}) will span only a part of the Hilbert space; this could be used to reduce the number of qubits significantly. Indeed, provided that we make the mapping between the two qubits to 1 qubit space:
\begin{eqnarray}
\{ |01 \rangle, |10\rangle\} \longrightarrow \{ | 0 \rangle , |1\rangle \}~{\rm or}~ \{ |00 \rangle, |11\rangle\} \longrightarrow \{ | 0 \rangle , |1\rangle \}   \nonumber 
\end{eqnarray} 
for the $S_z$ and parity symmetry cases respectively,  
then, both transformations (\ref{eq:01only})  and (\ref{eq:0011only}) become identical and equal to 
\begin{eqnarray} 
 U(\theta_1, \theta_2)  &=&
 \left[
  \begin{array}{cc}
\cos(\theta_1)  &e^{i\theta_2} \sin(\theta_1) \\
e^{-i\theta_2} \sin(\theta_1) & -\cos(\theta_1)   \\
% \cos( \Delta t \alpha )    &  0 & 0 & -i \sin(\sqrt{\Delta t \alpha})   \\
% 0     &  1 & 0 &  0   \\
% 0     &  0 & 1  &0  \\
% -i \sin(\sqrt{\Delta t \alpha})     & 0 & 0 &  \cos( \Delta t \alpha )
\end{array}
\right] \label{eq:reduced}
\end{eqnarray}
in the 1 qubit space. The circuit that performs the above unitary transformation is shown in Fig. \ref{fig:circsym2}-d.  In both cases, the qubit number to describe the system is divided by a factor $2$. Such technique has been explicitly used, for instance, in Ref. \cite{Cer21} where the occupation of lower and upper levels in a set of two-level systems was directly encoded as $|0 \rangle $ and $|1 \rangle$. 
A technique similar to the parity encoding was used, for instance, in Refs.~\cite{Kha21,Rui22} to encode the pair occupations instead of the particle occupations directly. 

\subsection{Examples of symmetry preserving quantum ans\"atze}

In a simple two qubits case, we have shown how the symmetry respected by a problem, similar to the classical computer 
case, can help reduce the resources required on a quantum computer. Below are a few illustrations of many-body interacting Fermi systems where symmetry-preserving ans\"atze have been employed.

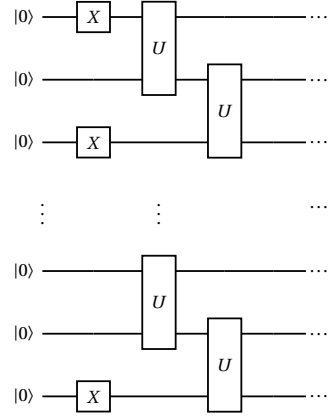
\begin{figure}[htbp]  
\centering
\begin{tikzpicture}
  \node[scale=0.85] { %scale=0.85
\begin{quantikz} 
%\qw & \qw & \qw \\
\lstick{ \ket{0}} & \gate{X} & \gate[wires=2]{U} & \qw               & \qw  & \ \ldots\ \qw & \\
\lstick{ \ket{0}} & \qw      &                   & \gate[wires=2]{U} & \qw  & \ \ldots\ \qw & \\
\lstick{ \ket{0}} & \gate{X} & \qw               &                   & \qw  & \ \ldots\ \qw & \\
\vdots            &          & \vdots            &                   &      & \ \ldots\     & \\
\lstick{ \ket{0}} & \qw      & \gate[wires=2]{U} & \qw               & \qw  & \ \ldots\ \qw & \\
\lstick{ \ket{0}} & \qw      &                   & \gate[wires=2]{U} & \qw  & \ \ldots\ \qw & \\
\lstick{ \ket{0}} & \gate{X} & \qw               &                   & \qw  & \ \ldots\ \qw & \\
\end{quantikz}
}; 
\end{tikzpicture}\\
    \caption{ Illustration of the construction of a wave-function with a fixed number of particles. The method consists in applying a set of layers 
    built from a sequence of 2-qubits unitary transformation between adjacent qubits. Each transformation takes the form (\ref{eq:01only}) and is associated with two parameters $(\theta_i, \varphi_i)$. In order to construct an efficient circuit in general a number $C^A_N$ of $U$ gates are needed where $N$ is the number of qubits and $A$ the number of particles Ref. \cite{Gar20}.}
    \label{fig:gard}
\end{figure}

\subsubsection{Particle number preserving states}
 
 In  Ref. \cite{Gar20}, a method that constructs states with a fixed number of particles was formulated. This technique generates an ansatz using the two-qubit gate shown in Eq. (\ref{eq:state01}). Using the Jordan-Wigner Transformation (JWT) mapping for fermions (see \ref{sec:fermion}), fixing the particle number is equivalent to retaining in the decomposition of the state given by Eq. (\ref{eq:wavegeneral}) only NB states where the number of 1 is precisely equal to the number of particles. As an alternative method, we mention the technique of Ref.~\cite{Aru20} to construct a variational Hartree-Fock (HF) state where the Thouless theorem \cite{Tho60,Rin80,Bla86} was used to create the quantum ansatz.
 
A generalization of the latter approach that allows us to include correlations beyond the HF approximation, is attracting increasing attention is the unitary coupled cluster (UCC) approach \cite{Rom19,Guo21}. This technique can eventually treat correlations beyond the HF theory. The UCC wave-function is written as:
\begin{eqnarray}
| \Psi(\bm{ \theta})\rangle &=& e^{\left[ T(\bm{ \theta}) - T^\dagger(\bm{ \theta}) \right]} | \phi \rangle . \label{eq:ucc}
\end{eqnarray}  
In practice, $| \phi \rangle$ is usually assumed to be a Slater determinant while  $T = T_1 + T_2 + \cdots$ is a set of multi-particle multi-hole (mp-mh) 
excitations truncated at a certain order, i.e. 
\begin{eqnarray}
\begin{array}{ll}
 T_1 = \sum_{p_1 h_1} \theta^{p_1}_{h_1} a^\dagger_{p_1} a_{h_1} ,      &{\rm (Single ~-~ 1p-1h)}    \\
 T_2 = \sum_{p_1 p_2 h_1 h_2} \theta_{p_1 p_2}^{h_1 h_2} a^\dagger_{p_1} a^\dagger_{p_2} a_{h_1}a_{h_2}   &{\rm (Double~-~2p-2h)} \\
 \vdots & \vdots
\end{array} \nonumber 
\end{eqnarray}  
One of the interesting aspects of this technique is that the particle number symmetry is automatically respected. In addition, circuits can be systematically obtained through the use of the Trotter-Suzuki transformation \cite{Tro59}. Detailed aspects of the UCC method can be found for instance in Ref.~\cite{Ana22}. It is worth noting that quantum computing applications to light atomic nuclei were often based on the coupled-cluster technique \cite{Kis22,Dum18}. 

\subsubsection{Total spin preserving states}

An illustration of problems where the spin symmetry is relevant is the Fermi-Hubbard model (see, for instance, the discussion in Refs.~\cite{Sek20,Sek22}). As we have seen previously in section \ref{sec:symcons}, eigenstates of ${\bf S}^2$  should respect specific properties concerning permutations; this renders the problem of constructing variational ansatz of this operator rather complex. In physics, the composition of spins and, more generally, angular momenta can be made using the Clebsch-Gordan transformation for which specific algorithms have been proposed \cite{Bac06,Kir17,Kir18,Hav18,Kro19}.  

A possible method to construct a specific eigenstate $| S, M \rangle$ of the total spin and its azimuthal projection on a quantum computer 
is to consider iteratively qubits and follow a specific path in the figure \ref{fig:spin}. Illustration of this methodology as well as the corresponding 
circuits can be found in Refs. \cite{Sug16,Sug19}. 
   
The quantum resources required to encode some physical systems might be significantly reduced when their Hamiltonian is block-diagonal, and each block has a given symmetry.
This was recently illustrated in Ref. \cite{Hla22} for solving the Lipkin-Meshkov-Glick (LMG) model \cite{Lip65}. In this model, a set of $A$
fermions can occupy a set of $A$ identical 2-level systems. A brute-force use of the Jordan-Wigner transformation to encode the problem would require one qubit per single-particle level, i.e., $2A$ qubits. Suppose one uses the fact that the total number of particles is conserved and that each 2-level system contains exactly one particle, one can encode the problem using quasi-spin algebra such that one encodes the occupation of the lower or upper level of the 2-level system directly into a single qubit. This technique was used in Ref.~\cite{Cer21} to reduce the number of qubits by a factor $2$. A more drastic reduction in the number of qubits can be achieved by using the fact that the system is fully symmetric with respect to the exchange of different two-level systems. One can then map the problem into a total spin problem and consider separate blocks with various eigenvalues $S$. Fully symmetric states, denoted by $|S,M \rangle$ correspond to the $2S+1$ states with only one line in the Young tableau (those with maximal $S$  values in Fig. \ref{fig:spin} with $S=A/2$). The Hamiltonian then becomes a tri-diagonal matrix in this basis and can be directly encoded, assuming that each $M$ is associated with a binary number. Then, the number of qubits $N$ to encode the problem is the minimal value for which $N> \ln(A+1)/\ln 2$. As shown in \cite{Hla22}, a further reduction in the circuit length was possible because odd and even values of $M$ are disconnected.
Altogether, for instance, for $A=1000$, the JWT method would require $2000$ qubits while taking advantage of all possible symmetries to encode the problem reduces the number of qubits to $10$. 

\subsection{Symmetry breaking quantum ans\"atze}
\label{sec:sbansatz}

Taking a quantum trial state that does not respect the symmetries of a physical problem 
might appear rather surprising in view of the discussion above. However, for some highly
 non-perturbative many-body problems, it is well known that the breaking of symmetries 
might be a very powerful and economical way to grasp correlations that might be extremely 
difficult to treat without breaking the symmetry. A typical example is the problem of superfluidity where 
quasi-particle vacuum are used that break the $U(1)$ symmetry associated to particle number symmetry. An
illustration of such state is the BCS ansatz that can be written in terms of creation/annihilation operators as \cite{Bri05}:
\begin{eqnarray}
| \Psi \rangle &=& \prod_{p>0} (u_p + v_p a^\dagger_p a^\dagger_{\bar p}) | - \rangle ,  \label{eq:BCS}
\end{eqnarray}   
where $(p,\bar p )$ refers to a pair of time-reversed state. We see that the state mixes different parts of the total Fock space having $0$, $2$, $4$,... number of particles.  
Such a state is known to be extremely useful to treat the long-range correlations between pairs of fermions while being rather easy to prepare 
both on classical and quantum computers compared to an efficient symmetry-preserving ansatz \cite{Ver09,Jia18,Lac20}. The BCS ansatz can be written in the qubit basis as \cite{Lac20,Kha22}:
\begin{eqnarray}
 | \psi \rangle = \prod_{n}  \left[ \cos \left( \frac{\theta_n}{2}\right) I_{n} \otimes I_{n+1} + \sin\left( \frac{\theta_n}{2}\right) Q^+_n Q^+_{n+1} \right] | - \rangle, \label{eq:bogo2} \nonumber
\end{eqnarray} 
where $(n, n+1)$ labels the qubit  numbers associated to the particles $(p, \bar p)$. Here we have made the replacement $u_p = \cos (\theta_n /2)$ 
and  $v_p= \sin (\theta_n /2)$ using the fact that $u_p^2+v_p^2=1$. If we focus on a single pair, we recognize a generalization of a Bell state 
$[\cos \left(\theta /2 \right) | 00 \rangle +\sin\left(\theta / 2\right)| 11 \rangle]$ similar to the one given in Eq. (\ref{eq:genbell}).  If then we consider a system with even number of particles and no pair is broken (seniority zero), 
we can further reduce quantum resources by directly encoding the pair creation operator $P^\dagger_{p} = a^\dagger_p a^\dagger_{\bar p}$ using the JWT.  This method was used with success to describe simple superfluid systems, for instance, in Refs.~\cite{Kha21,Rui22,Rui21}.  Considering only one pair, the associated generalized Bell state is simply simulated as a rotation on a single qubit, i.e. 
\begin{eqnarray}
R_Y(\theta) |0\rangle_p = \cos \left(\theta /2 \right) | 0 \rangle_p +\sin\left(\theta / 2\right)| 1 \rangle_p. \nonumber 
\end{eqnarray} This simple example illustrates that correlations in superfluid systems can be grasped using a set of “independent” rotations on qubits, where each qubit describes
a pair of particles.  As an alternative to the BCS approach, one can eventually impose precisely the number of particles and use the construction of quantum ansatz depicted in Fig.~\ref{fig:gard}; however, the number of operations to properly treat correlations would be much higher.

A critical aspect of the use of a symmetry-breaking state is that a precise comparison with observations usually requires the restoration of symmetries that have been broken in the first step. Such symmetry breaking-symmetry restoration strategy is currently widely used, for instance, in the context of atomic nuclei \cite{Rin80,Bla86,Ben03,Rob18,She19}. The discussion below will concentrate on the definition of projectors and symmetry restoration. 

\subsubsection{Unwanted symmetry breaking}

Before discussing symmetry restoration on quantum computers, we should discuss as well the problem of symmetry breaking in the context of using noisy devices. Even if a symmetry-preserving state is constructed, due to the limited fidelity of each operation, it is improbable that the imposed symmetry would be perfectly respected when running the circuit. Unwanted breaking symmetries might occur both during the construction or the manipulation of the state. For an illustrative example, let us assume a state that decomposes solely on NB states with a fixed number of qubits in state $1$ (i.e., fixed particle number in the many-body context). Typical examples of errors that will break this property are the so-called spin-flip or relaxation process that leads to unwanted jumps between $0$s and $1$s, or the cross-talks effects between qubits. Because of this, an unwanted admixture of symmetry-breaking components will occur during the circuit processing. Respecting the symmetry or not might be a method to control the errors \cite{Saw16,McA19}. Different methods have been proposed to perform error-mitigation that are based on symmetries. To quote some of them, we mention the symmetry protection \cite{Tra21} , symmetry verification \cite{Got97,Bon18,Sag19,Tra21} and/or symmetry distillation \cite{Koc21,Hug21}.

\section{Symmetry restoration by projection and oracles}
\label{sec:oracle}

We discuss the possibility of using the concept of oracles for symmetry restoration. Oracles are essential for historical quantum search algorithms like the Grover algorithm \cite{Gro97a,Gro97b}. As far as we know, this possibility has not been explored for symmetry restoration. We start below with projectors, the standard way in classical computers to perform symmetry restoration, and then we show how these projectors can be connected to oracles. The oracle will then be used in the amplitude amplification method proposed in section \ref{sec:amplitude}. 

\subsection{General discussion} 

As we mentioned previously, symmetries are associated to partition the entire Hilbert space ${\cal S}$ into a set of 
sub-spaces, denoted generically as $\{ {\cal S}_\alpha \}$. These sub-spaces are not connected by the Hamiltonian of the system.
Having a general wave-function $| \Psi \rangle$ written in the full space, a common and natural way to restrict the 
wave-function to one of the sub-spaces is to use the projector $\hat P_\alpha$ onto the subspace of interest. This projector verifies $\hat P_\alpha^2 = \hat P_\alpha$. For the symmetry problem, we usually consider the case ${\cal S}=\bigcup_\alpha {\cal S}_\alpha$, such that we have the closure relation:
\begin{eqnarray}
\sum_\alpha \hat P_\alpha = I. 
\end{eqnarray}  
The probability 
amplitude of the wave-function to belong to ${\cal S}_\alpha$ is given by
\begin{eqnarray}
p_\alpha (\Psi) &=& \langle \Psi | \hat P_\alpha | \Psi \rangle , 
\end{eqnarray}  
while the corresponding normalized projected wave-function writes
\begin{eqnarray}
| \Phi_\alpha  \rangle &=& \frac{1}{\sqrt{p_\alpha(\Psi)}}  \hat P_\alpha |  \Psi \rangle.  \label{eq:psiproj} 
 \end{eqnarray}
We consider the specific situation where each sub-space ${\cal S}_\alpha$ corresponds to 
 the set of wave functions that respect a certain symmetry associated with a symmetry operator $\hat S$. Specifically, we 
 restrict the discussion to cases where the operator $\hat S$ can take a limited set of discrete values 
 $\{ \lambda_\alpha \}_{\alpha=1,M}$ which have finite lower and upper limits. All cases presented in section \ref{sec:symcons}  
 enter into this class of symmetries. In practice, the subspace ${\cal S}_\alpha$ is associated with the set of eigenstates 
 of $\hat S$ having the specific eigenvalue $\lambda_\alpha$.      
 
 Often in physics and chemistry, several symmetries are respected simultaneously. The above consideration can be simply generalized by considering 
 a set of operators $\{ \hat S_1,   \hat S_2 , \cdots ,\hat{S}_\Gamma \}$ where each operator corresponds to one symmetry.  Provided that the symmetries are compatible, i.e., we have $[\hat S_\gamma, \hat S_\delta ] =0$ with $\gamma$ or $\delta = 1, \cdots \Gamma$, one can define the sub-spaces ${S_{\alpha_1, \cdots , \alpha_\gamma}}$ formed of states respecting all symmetries simultaneously and associated with the projectors 
 $\hat P_{\alpha_1, \cdots , \alpha_\Gamma} = \hat P_{\alpha_1} \cdots  \hat P_{\alpha_\Gamma}$. For most of the discussions below, we focus on one symmetry, keeping in mind that the discussions can be generalized easily to several symmetries. 

\subsection{Various forms of projectors}

In the context of quantum computing, it is important to recall some properties of projectors. The first one is that the projectors are hermitian operators but 
are not unitary. Since perfect quantum computers can only perform unitary operations, one cannot directly encode the projector on a quantum computer by performing only unitary operations on the system. A second property is that the projectors can be expressed in different ways. While the non-unitarity can be seen more like a difficulty for projection on QC, one can take advantage of the second property to design multiple projection techniques \cite{Yen19,Izm19} or reduce the quantum resources needed for the projection.

To illustrate the different types of expressions one can use on a quantum computer, we consider the projector $\hat P_\alpha$ that projects onto the subspace associated with the eigenvalues $\lambda_\alpha$. A compact expression for the projector is given by
\begin{eqnarray}
\hat P_\alpha &=& \delta (\hat S - \lambda_\alpha). \label{eq:projdelta} 
\end{eqnarray}  
Such form, that is useful for formal manipulation, is never used when trying to implement the projector numerically on whatever type of
computer architecture (quantum or classical).       
  
% here the case where the symmetry operator 
%$\hat S$ identifies with the particle number $\hat N$. 
%We consider below the projector that project onto the subspace of wave-functions with exactly 
%$N$ particles and denote by $\hat P_N$ the associated projector. Illustrations of physical situation where such projection was performed in QC can be found in Refs. \cite{Lac20,Kha21,Rui21,Rui22,Kha22}.  

A possible alternative for the projector is to use the formula \cite{Low55,Low69}:
\begin{eqnarray}
\hat P_\alpha &=& \prod_{\beta \neq \alpha} \left[  \frac{\hat S - \lambda_\beta}{\lambda_\alpha - \lambda_\beta} \right]. \label{eq:projlow}
\end{eqnarray} 
Such equation has been used, for instance, in Ref.~\cite{Mol16} to obtain a compact expression of the Hamiltonian iteratively in a reduced Hilbert space. In the case of parity projection, which has only two eigenvalues $\pi = \pm 1$, one can rewrite this projector simply as
\begin{eqnarray}
\hat P_\pi &=& \frac{1}{2} \left( 1 + \pi \hat S\right). \label{eq:piproj}
\end{eqnarray}
For this simple situation, the projector can be employed directly as a practical tool to verify or enforce symmetries \cite{Bon18,Sag19}.   
For other symmetries, the expression (\ref{eq:projlow}) turns out to be rather complex to manipulate. Indeed, for instance, for the 
particle number of spin operators, $\hat S$ is a one- or two-body operator; therefore, the different powers $\hat S^k$ that appear when developing Eq. (\ref{eq:projlow}) are k-body or 2k-body operators that are generally difficult to implement.     

To discuss an alternative form of the projector, we first consider the case where the symmetry operator $\hat S$ identifies with the particle number $\hat N$. We denote by $\hat P_N$ the projector onto the subspace of wave functions with exactly $N$ particles. Illustrations of the physical situation where such projection was performed in QC can be found in Refs. \cite{Lac20,Kha21,Rui21,Rui22,Kha22}.  A standard method used on classical computers consists in writing the projector as an integral \cite{Rin80,Bla86,Ben03,Rob18,She19} \footnote{The equivalence between the expression (\ref{eq:projdelta}) and this expression 
can be shown by noting simply that: 
\begin{eqnarray}
\int_{0}^{2\pi}   \frac{d\varphi }{2\pi} e^{i(N - k) \varphi} &=& \delta(N-k) \nonumber 
\end{eqnarray} 
}:
\begin{eqnarray}
\hat P_N  &=& \int_0^{2 \pi} \frac{d\varphi} {2\pi} e^{i\varphi (\hat N - N) } . \label{eq:projgauge}
\end{eqnarray}
Here $\varphi$ plays the role of the conjugated variable of the particle number and is usually called gauge angle.  
The integral on gauge angle can be efficiently discretized on a classical device using the  Fomenko method \cite{Fom70,Ben09}. 

Finally, as an alternative to the integral form given by Eq. (\ref{eq:projgauge}), one can directly use a discrete representation of the 
$\delta$ function provided by:
\begin{eqnarray}
\hat P_N &=& \frac{1}{M+1} \sum_{k=0}^{M} e^{2 \pi i k (\hat N- N)/(M+1)} , \label{eq:pngen4}
\end{eqnarray}
where $M$ denotes the maximum number of particles allowed in the Hilbert space. When a many-body problem is considered, and the JWT is used, $M$ identifies with the number of qubits. The expression (\ref{eq:pngen4}) is particularly interesting in the QC context since each term defined by $U_k (N) = e^{2 \pi i k \hat N/(M+1)}$ corresponds to a unitary operation. Accordingly, the operator $\hat P_N$ is written as a linear combination of unitaries (LCU) operators for which specific quantum computing algorithms have been proposed \cite{Chi12,Ber14,Ber15}. It should be noted that the discretized form of Eq. (\ref{eq:projgauge}) can also lead to an approximation of the projector with various precisions depending on the number of points used to compute the integral. 

A similar discussion can be made for other symmetries. 
For instance, for the angular momentum projection, the projector can be expressed as integrals of different orientations through the Wigner method \cite{Rin80}. The corresponding reduction to an LCU was discussed for QC in Refs. \cite{Sek20,Sek22,Tsu20,Tsu22}. In the following, we will use the generic LCU form for any projector:
\begin{eqnarray}
\hat P_\alpha &=& \sum_l^{\Lambda} \beta^{(\alpha)}_l \hat V^{(\alpha)}_l \label{eq:LCU}   
\end{eqnarray} 
where $\beta^{(\alpha)}_l$ and $\hat V^{(\alpha)}_l$ are a set of constants and unitary operators respectively.  

Coming back to the particle number projection, we finally quote a last possible expression for the projector that we will use below \cite{Ata73,Sme73}:
 \begin{eqnarray}
\hat P_N &=& \prod_{l=0}^{k} \frac{1}{2} \left( 1 + e^{i\phi_l (\hat N - N)}\right) \label{eq:prodN}
 \end{eqnarray}
 with $\phi_l = \pi/2^{l}$ with $k= [\log_2 {\rm max} \{N, M-N \}]$ was discussed and/or used to perform projection on QC recently in 
 Ref. \cite{Yen19,Kha21}. We show in \ref{app:proofsprojN} the strict equivalence between (\ref{eq:pngen4}) and (\ref{eq:prodN}).

\subsection{From projectors to oracles}
\label{subsec:oracle}

Oracles are ``magical" operators that are the basis of some of the historical quantum search algorithms such as the Grover method \cite{Gro97a,Gro97b}. A complete discussion of the Grover and more general search algorithms is out of the scope of the present article, so, for further discussions on the subject, we recommend the textbooks \cite{Kay11,Lim19}. 

We focus here on constructing an oracle from a projector after giving below a simplified discussion on its definition. The basic assumption of the oracle is that the total Hilbert space can be separated into two complementary subspaces ${\cal H}_{G}$ and  ${\cal H}_{B}$. These subspaces contain all states having (``good states") or not (``bad states") a specific property. The oracle is a unitary operator, often denoted by $U_f$ having the following property: any state in ${\cal H}_{G}$ (resp. ${\cal H}_{B}$) will be an eigenstate of $U_f$ with eigenvalue $-1$ (resp. $+1$). A general state of a system is a priori a superposition of good and bad components. The oracle is one of the critical components that can help reduce the contribution of ``bad" elements in favor of ``good" ones. 

\begin{figure}
\begin{center}
 \includegraphics[width=\columnwidth] {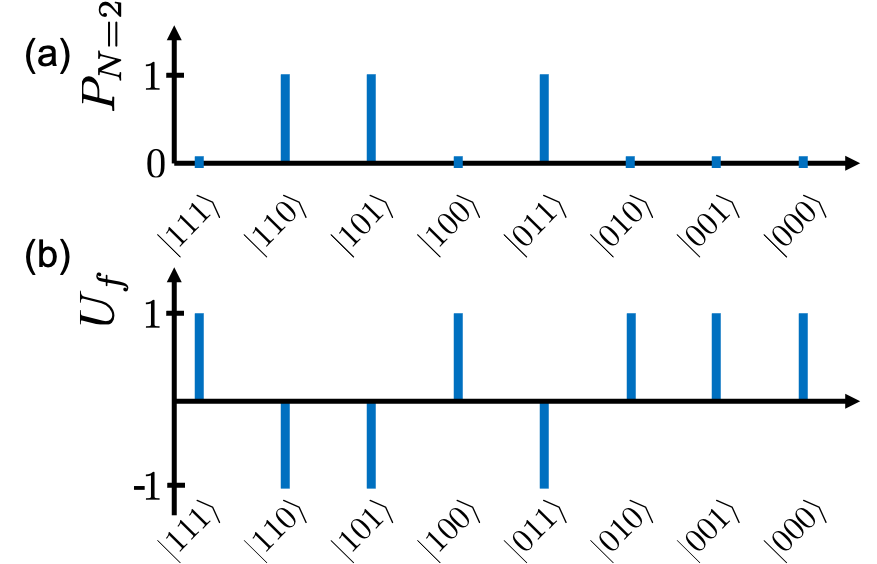}  %width=0.9\linewidth
 \end{center}
\caption{Panel a: Illustration of the eigenvalues of the projector on $A=2$ particles in a system formed of 3 single-particle states encoded on 
3 qubits using the JWT. In this case, all states of the NB are eigenstates of the projector with eigenvalues $0$ or $1$. The 
eigenvalues of the associated oracle are shown in panel b.}
\label{fig:oracle}       
\end{figure}

In symmetry restoration, one might see the ``good" components as those that respect a certain symmetry, while ``bad" components are those that do not respect it. Starting from a projector defined above, one can efficiently get the oracle having the property above. Let us consider the unitary operator 
\begin{eqnarray}
\hat U(\omega) = e^{i \omega \hat P_\alpha}
\end{eqnarray}
where $\hat P_\alpha$ is one of the projectors that restores a symmetry.
Using the properties of projectors, we obtain:
\begin{eqnarray}
\hat U(\omega)  &=& I + \left( e^{i\omega} - 1 \right) \hat P_\alpha. \nonumber 
\end{eqnarray}  
The subspace ${\cal H}_{G}$ contains all the states that respect the symmetry. Therefore, this space identifies with ${\cal S}_\alpha$. We define the projector on the rest of the Hilbert space, i.e., on ${\cal H}_{B}$ by $\hat Q$ (note that $ \hat P_\alpha + \hat Q = I$). Any state in ${\cal S}_\alpha$  (resp. ${\cal H}_{B}$) is an eigenstate of $\hat U(\omega)$ with eigenvalue  $e^{i \omega}$ (resp. $1$).  We therefore see that the standard oracle is obtained simply by taking $\omega = \pi$, i.e. $\hat U_f = e^{i \pi \hat P_\alpha}$. Such oracle can be implemented in practice on QC, for instance, using the Trotter-Suzuki decomposition \cite{Tro59,Bab15,McA20}. We illustrate in Fig. \ref{fig:oracle} the effect of a projector and oracle in the NB for the case of the particle number symmetry.  The explicit use of the oracle for symmetry problems will be discussed in section \ref{sec:amplitude} .

%For instance, for systems with good number parity, the integral can be reduced to $[0,\pi]$ and discretized as    

\section{Restoration of broken symmetries on quantum computer}    
\label{sec:bs}

In previous sections, we discussed the different facets associated with symmetries in a quantum problem. 
We have, in particular, introduced some of the operators of the most common symmetries in many-body problems. In classical computers, projectors are handy tools that can directly be used to restore symmetries. In a quantum computer, the projection becomes more delicate because projectors are non-unitary operators that are not easy to implement on a QC. Below we describe different methods that can be used to restore symmetries in the QC context. Some of the ways below have already been used previously, while others, as far as we know, are new. The presented methods do not have all the same targeted goals. Some of them will give access to expectation values of observables without requiring the construction of the projected state. In contrast, others will produce the projected state as an outcome of the circuit that could be used for post-processing. As an obvious consequence of the differences in goals, quantum resources required to restore symmetries will also vary, which is an essential aspect of the NISQ period. Note that we tested and validated most of the methods below using the Qiskit toolkit \cite{Abr19}. 

\subsection{Observation of projected state with classical post-processing: A hybrid approach}

Performing explicit projection with a quantum computer often requires to consider additional register qubits, carrying out a large set of operations, or both (see discussion below). When the projected state is not explicitly needed, but only the expected values of observables with this state, complexity can be reduced by using a hybrid approach where a classical computer performs parts of the tasks. We illustrate two of the classical post-processing methods giving access to the symmetry-restored state processing. 

\subsubsection{Projection based on Hadamard test}
Let us consider a state $| \Psi(\bm{ \theta}) \rangle$ that is prepared on a QC. This state is supposed to break the symmetry that could be restored with the projector $\hat P_\alpha$. The aim is to compute the observable $\hat O$; a typical example of observable is the Hamiltonian itself. We will assume that $\hat O$ commutes with $\hat P_\alpha$.  Then, the expectation value on the projected state 
(\ref{eq:psiproj}) can be recast as:
\begin{eqnarray}
\langle \hat O \rangle &=& \frac{\langle \Psi(\bm{ \theta}) | \hat O \hat P_\alpha | \Psi(\bm{ \theta}) \rangle }{\langle \Psi(\bm{ \theta}) | \hat P_\alpha | \Psi(\bm{ \theta}) \rangle} = \frac{ \sum_{l} \beta_l \langle \hat O \hat V_l  \rangle_{\bm{ \theta}} }{ \sum_{l} \beta_l \langle  \hat V_l  \rangle_{\bm{ \theta}}}, 
\label{eq:oexp}
\end{eqnarray}     
where, on the left-hand side, we used Eq. (\ref{eq:LCU}) omitting the label $^{(\alpha)}$, and where we introduced the notations 
$\langle . \rangle_{\bm{ \theta}} = \langle \Psi(\bm{ \theta}) | . | \Psi(\bm{ \theta}) \rangle$ for compactness.  The $\hat V_l $  are independent unitary operators that usually can be implemented directly on the QC. Provided that the operator $\hat O$ itself can be implemented by a circuit, the set of expectation values $\langle \hat O \hat V_l  \rangle_{\bm{ \theta}}$ and  $\langle  \hat V_l  \rangle_{\bm{ \theta}}$ can be obtained by a set of independent Hadamard tests \cite{Nie02,Kay11,Lim19}. Then, the different expectation values obtained by a set of independent measurements can be collected 
and the Eq. (\ref{eq:oexp}) is performed a posteriori on a classical computer.  This pragmatic approach was used for many-body systems in Ref. 
\cite{Kha21} and for  error mitigation in Ref.  \cite{Bon18,Sag19}.  

\subsubsection{Projection based on LCU method}
A generalization of the above method, when the operator $\hat O$ itself is hermitian but not unitary, is to write it as a linear combination $\hat O = \sum_k w_k \hat W_k$ of unitaries $\hat W_k$. Then, we need to obtain the enlarged set of expectation values $\langle \hat W_k \hat V_l  \rangle_{\bm{ \theta}}$ and  $\langle  \hat V_l  \rangle_{\bm{ \theta}}$. Note however that for complex systems the number of operators grows rapidly when the number of particles (qubits) increases.    
 
Guided by the integral form of the projector (Eq.~\eqref{eq:projgauge}), one can also propose a different type of classical post - processing based on the concept of generating function. Generating functions were recently used in Ref. \cite{Rui21} to obtain the spectroscopy of many-body states and can also be employed to perform the projection. Using Eq.~\eqref{eq:projgauge}, one can rewrite Eq. \eqref{eq:oexp} as:
\begin{eqnarray}
\langle \hat O \rangle &=& \frac{ \displaystyle \int_0^{2 \pi}  d\varphi e^{-i \varphi N} \langle \hat O e^{i\varphi \hat N}\rangle_{\bm{ \theta}}}
{\displaystyle \int_0^{2 \pi}  d\varphi  e^{-i \varphi N}  \langle e^{i\varphi \hat N}\rangle_{\bm{ \theta}}  }.
\label{eq:gf}
\end{eqnarray}     
Similarly to the previous method, the expectation values $ \langle \hat O e^{i\varphi \hat N}\rangle_{\bm{ \theta}}$ and $\langle e^{i\varphi \hat N}\rangle_{\bm{ \theta}}$ for different $\varphi \in [0, 2 \pi]$ can be evaluated on the QC while the integral is performed on the classical device. A similar treatment can be made for the angular momentum projection \cite{Sek22,Tsu20,Tsu22}. 

In this section, we discussed methods that do not require the projected state as output. Below, we consider different possible strategies to obtain also the symmetry-restored states.

\subsection{Implementing projection on a quantum computer}
\label{sec:implementation}

%\textcolor{ForestGreen}{I think most of the sections onward including this one are part of section 5. So, they can be subsections of section 5.}

A standard limitation in QC is that it is bound to perform unitary operations. As mentioned above, the projection operators are hermitian but not unitary. Here we discuss possible methods that 
leads to the procedure:
\begin{eqnarray}
| \Psi \rangle \rightarrow \hat P_\alpha |  \Psi \rangle, \label{eq:proj} 
\end{eqnarray}
where $\hat P_\alpha$ is one of the projectors associated with symmetry restoration. In some cases, the method discussed below can apply to any projection onto a given subspace of the total Hilbert space. The solution to the problem of Eq. (\ref{eq:proj}) usually requires the addition of a set of ancillary qubits and the transfer of information for the system register to this additional set of qubits. The projection is then performed by measuring these extra qubits (indirect measurement). Three different methods will be presented below: the LCU, the QPE (Quantum Phase Estimate), and the iterative IQPE-like approach \cite{Nie02}.   In addition to these methods, we will present an alternative approach based on the Grover search algorithm \cite{Gro97a,Gro97b} that, a priori, does not require the addition of ancillary qubits.

\subsubsection{Projection using the LCU technique}
\label{sec:lcu}

Provided that the operator $\hat P_\alpha$ can be written as a linear combination of unitary operators as given in Eq.~\eqref{eq:LCU}, and that each of these operators can be efficiently implemented in a circuit, a possible way to perform the projection is the LCU method proposed in Refs. \cite{Lon06,Chi12,Ber14,Ber15}. 
The LCU method is a generic approach to apply a non-unitary operator on a state encoded on a quantum register. We show here that this method can be a valuable tool in the post-NISQ era to perform symmetry restoration.

%(c) \\
%\begin{tikzpicture}
%  \node[scale=0.75] {
%\begin{quantikz} 
% %\lstick{ \ket{0}}  & 
% \push{} &  \ctrl{1} & \gate{X} &\gate{R^{\dagger}(\theta_1, \theta_2)}  & \targ{}  & \gate{R(\theta_1, \theta_2)}  & \gate{X}   & \ctrl{1} &   \push{}  \qw \\
% %\lstick{\ket{0}}  & 
% \push{}  &\targ{}  & \push{} & \push{} &\octrl{-1} &\push{} &\push{} &\targ{} &  \push{} \qw 
%\end{quantikz}
%%\end{tikzpicture} \\

\begin{figure}[htbp]  
\centering
\begin{tikzpicture}
  \node[scale=0.8] { %scale=0.7
\begin{quantikz} 
%\qw & \qw & \qw \\
\lstick[wires=4]{$n_{LCU}$} &\lstick{ \ket{0}}    & \gate[wires=4, nwires={3}]{B} & \octrl{1}           & \ctrl{1}           & \ \ldots\ \qw & \qw               & \ctrl{1}              & \gate[wires=4, nwires={3}]{B^{\dagger}} & \meter{} & \\
                      &\lstick{ \ket{0}}    &                               & \octrl{1}           & \octrl{1}          & \ \ldots\ \qw &                   & \ctrl{1}              &                                         & \meter{} & \\
                      &\vdots               &                               & \vdots              & \vdots             & \ddots        &                   & \vdots                &                                         & \vdots   & \\
                      &\lstick{ \ket{0}}    &                               & \octrl{1} \vqw{-1}  & \octrl{1} \vqw{-1} & \ \ldots\ \qw & \qw               & \ctrl{1} \vqw{-1}     &                                         & \meter{} & \\
                      &\lstick{ \ket{\psi}} & \qw                           & \gate{U_{0}}        & \gate{U_{1}}       & \ \ldots\ \qw & \qw               & \gate{U_{2^{n_{LCU}}- 1}}  & \qw                                     & \qw      & \\
\end{quantikz}
}; 
\end{tikzpicture}\\
    \caption{ LCU circuit for $n_{LCU}$ ancilla qubits. This circuit can implement the linear combination of up to $2^{n_{LCU}} $ unitary matrices. We recall that the filled circles are the standard controlled operation by the $| 1 \rangle$ state, while the open circle is the controlled operation by the $| 0 \rangle$ state. The present sequence of operations scans all possible values of $l$ written as binary numbers from $0$ to $\Lambda$.}
\label{fig:projLCU}
\end{figure}
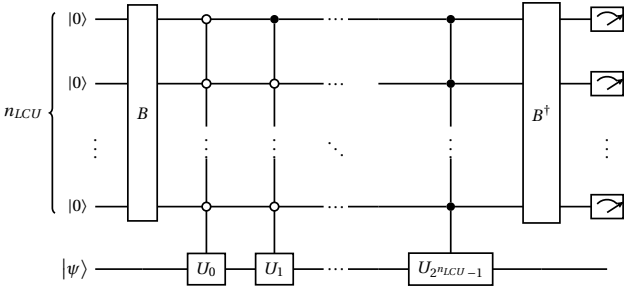

Our starting point is again to assume that we can decompose the projector as in Eq. (\ref{eq:LCU}). We assume that the number of unitary operators $\Lambda$ in this equation is finite and simply write $\beta^{(\alpha)}_l= \beta_l$. We also suppose that $\beta_l \ge 0$ for all $l$; this is always possible since a minus sign can be incorporated in the unitary operator simply through the replacement $\hat V_l \rightarrow - \hat V_l$. The system $|\psi\rangle$ is assumed to be described on $N$ qubits. We add to these qubits $n_{LCU}$ ancilla qubits to implement the LCU technique.

The LCU method consists of three main steps. The first step of the method is to prepare a state in the ancillary register such that:
\begin{eqnarray}
| B \rangle &=& \sum_{l=0}^\Lambda \gamma_l | l \rangle.  \label{eq:LCUregister}
\end{eqnarray} 
Here, we define the coefficients $\gamma_l$ as $\gamma_l = \sqrt{\beta_l} / \sqrt{\sum_l \beta_l}$. The states $|l\rangle$ are the natural basis states of the $n_{\rm LCU}$ qubits. We introduce the operator $\hat B$ such  that $| B \rangle = \hat B |- \rangle_{n_{\rm LCU}}$ (see Fig. \ref{fig:projLCU}).  
For the decomposition (\ref{eq:LCUregister}) to be possible we have the constraint $2^{n_{\rm LCU}} -1 \ge \Lambda$. We also assume 
that $\beta_l=0$ if $l > \Lambda$. The implementation of $\hat B$ can be made using for instance an optimized version of the quantum
Shannon decomposition (QSD) algorithm \cite{She06,Abr19}. Another method to prepare the state $\ket{B}$ is presented in Ref.~\cite{Siw02}.

The second step consists of a sequence of controlled operations correlating the system and the ancillary register. 
Each step in the sequence is associated with one of the $\hat V_l$, and the control operation is performed by the $| l \rangle$ state 
such that it leads schematically to the replacement $| \Psi \rangle \otimes | l \rangle \rightarrow \hat V_l | \Psi \rangle \otimes | l \rangle$ while other components are untouched.  An illustration of the circuit performing such operation is shown in Fig. \ref{fig:projLCU}.  After the sequence of operations, we end up with the state $\sum_l \gamma_l \hat V_l | \Psi \rangle \otimes | l \rangle$. 

% \begin{figure}[htbp]  
% \centering
% \\
% \begin{tikzpicture}
%   \node[scale=0.75] {
% \begin{quantikz}
% & \octrl{1}  & \qw \\
% & \gate{H}   & \qw
% \end{quantikz}
% =\begin{quantikz}
% \qw & \gate{X} & \ctrl{1} & \gate{X} & \qw \\
% \qw & \qw      & \gate{H} & \qw      & \qw
% \end{quantikz}
% }; 
% \end{tikzpicture}\\

% %\includegraphics[scale=0.3,angle=-90]{} 
%     \caption{Equivalence between a Hadamard gate controlled by the state $|0\rangle$\emph{
% (Left)} and a Hadamard controlled by the state $|1\rangle$ plus additional $X$
% gates\emph{ (Right)}.\label{fig:Equivalence-between_controls}. \textcolor{blue}{(Denis: again figure to be redone. I am not sure we should keep it, maybe it is too simple.)} }
% \label{fig:control0}
% \end{figure}

The last step of the method is to apply $\hat B^\dagger$ on the ancillary register. After this final step, it could be shown  that the total state decomposes as:
\begin{eqnarray}
| \Psi_{tot} \rangle &=& \left( \sum_l|\gamma_l |^2 \hat V_l | \Psi\rangle \right) \otimes |- \rangle_{n_{\rm LCU}} + \cdots  
\end{eqnarray} 
Therefore, if we measure only $0$s in the set of ancillary qubits for all $n_{\rm LCU}$ qubits, we 
automatically get that the system wave-function after measurement identifies with:
\begin{eqnarray}
\sum_l |\gamma_l|^2 \hat V_l | \Psi \rangle &=& {\cal C} \sum_l \beta_l  \hat V_l | \Psi \rangle  ={\cal C}  \hat P_\alpha | \Psi \rangle,
\end{eqnarray} 
which is nothing but the projected component of the initial state. Here ${\cal C}$ is a normalization constant that properly ensures a normalized state as an outcome of the measurement.

\subsection{Projection by amplitude amplification }
\label{sec:amplitude}

Most of the methods proposed previously to perform projection require one or several ancillary qubits together with the rejections of some events. We present here a new methodology in the context of projections that do not require any additional qubit or the need to reject an event. 
The method is inspired by the search algorithm of Grover \cite{Gro97a,Gro97b} and uses the oracles discussed in section 
\ref{sec:oracle} that are associated with projectors associated with symmetry restoration.

\begin{figure}
\begin{centering}
\includegraphics[width=0.7\linewidth] {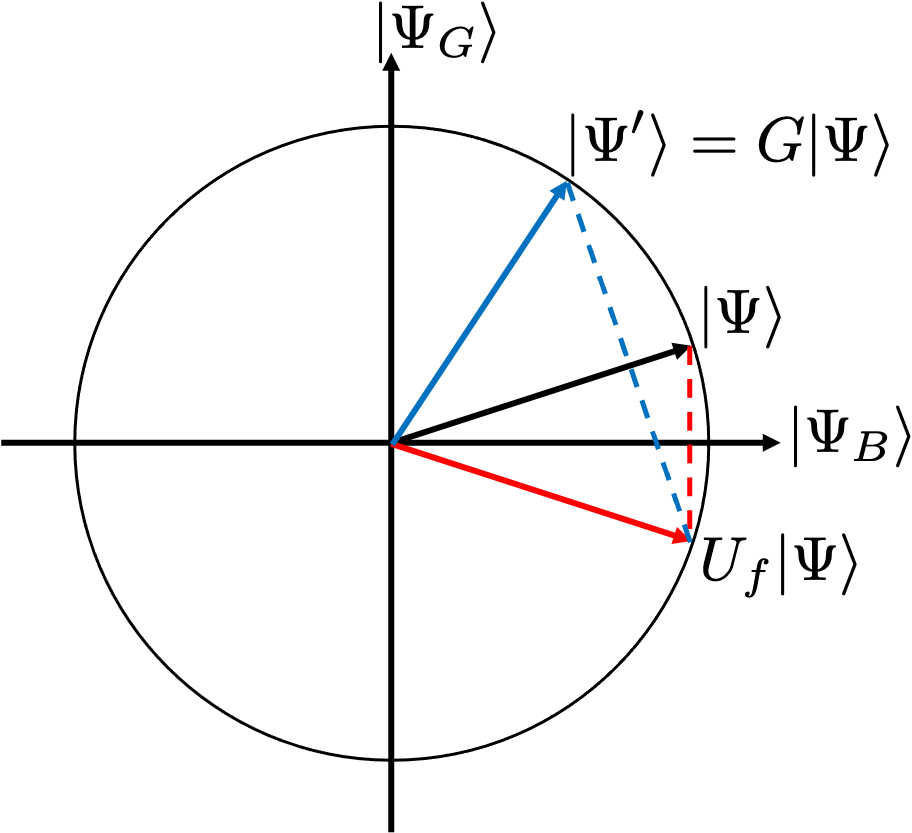} %width=\columnwidth
\par\end{centering}
\caption{Illustration of a single step of the Grover method applying the oracle operator $\hat U_f$ and the operation $\hat R_\Psi$ sequentially starting from the initial state $|\Psi \rangle$. In this example, the final state $| \Psi' \rangle = \hat G | \Psi \rangle$ verifies $|\langle \Psi' | \Psi_G\rangle |^2 > |\langle \Psi | \Psi_G\rangle |^2 $. }
\label{fig:grover}
\end{figure}

We now consider the only method where a priori no ancillary qubits should be added to perform the projection. Our starting point is the standard Grover search algorithm \cite{Gro97a,Gro97b} or, more generally, the amplitude amplification technique. This algorithm, recognized as one of the few that present quantum speed-up, is well documented. We refer to textbooks for a complete discussion \cite{Kay11,Lim19} and  focus here on a schematic view of the method's primary goal. Let us assume that we start from a state written as:  
\begin{eqnarray}
| \Psi \rangle &=& \alpha_0 | \Psi_G \rangle  + \beta_0   | \Psi_B \rangle.  \label{eq:psi}
\end{eqnarray}  
The state $| \Psi_G \rangle $ and $| \Psi_B \rangle $  ($G/B$ for good/bad) are respectively the states we search for or want to eliminate.
The essence of the Grover or amplitude amplification method is to iteratively increase (resp. decrease) the contribution of the good state (resp. the bad state). For this, two operations are used:
\begin{enumerate}
  \item An oracle operator $\hat U_f$ applied to any $|k\rangle$ state of the computational basis is defined as: 
  \begin{eqnarray}
    \hat U_f |k\rangle = (-1)^f |k\rangle \text{ with }
         \begin{cases}
           f=0, &\quad\text{if } |k\rangle \in |\Psi_B \rangle\\
           f=1, &\quad\text{if } |k\rangle \in |\Psi_G \rangle\\
         \end{cases}.
    \label{eq:def_oracle}
    \end{eqnarray}
We already discussed in section \ref{sec:oracle} how such oracle can be constructed from a projection operator. We also illustrated in Fig. \ref{fig:oracle} the effect of an oracle.  
  \item An operator $R_\Psi$ is associated to the input state $| \Psi\rangle$ 
  \begin{eqnarray}
\hat R_{\Psi} = 2 | \Psi \rangle \langle \Psi | - I.
\end{eqnarray}
This second operation, shown in the Fig. \ref{fig:grover} in blue, can be seen as a vector reflection with respect to the initial state.
\end{enumerate} 
We introduce the Grover operator $\hat G= \hat R_\Psi \hat U_f$ that combines the two operations. 
A schematic illustration of the two operations starting from the state $| \Psi \rangle$ is shown in Fig. \ref{fig:grover}.  
The repeated application of the $\hat G$ unitary transformation will induce a set of state rotations along the circle shown in Fig.~\ref{fig:grover}. Accordingly, the amplitude $p_n(\Psi)$, 
defined as 
\begin{eqnarray}
p_n (\Psi) &=& |\langle \Psi_G |\hat G^n| \Psi \rangle |^2,
\end{eqnarray}
varies as $n$ increases. 

\begin{figure}
\begin{centering}
\includegraphics[width=\columnwidth] {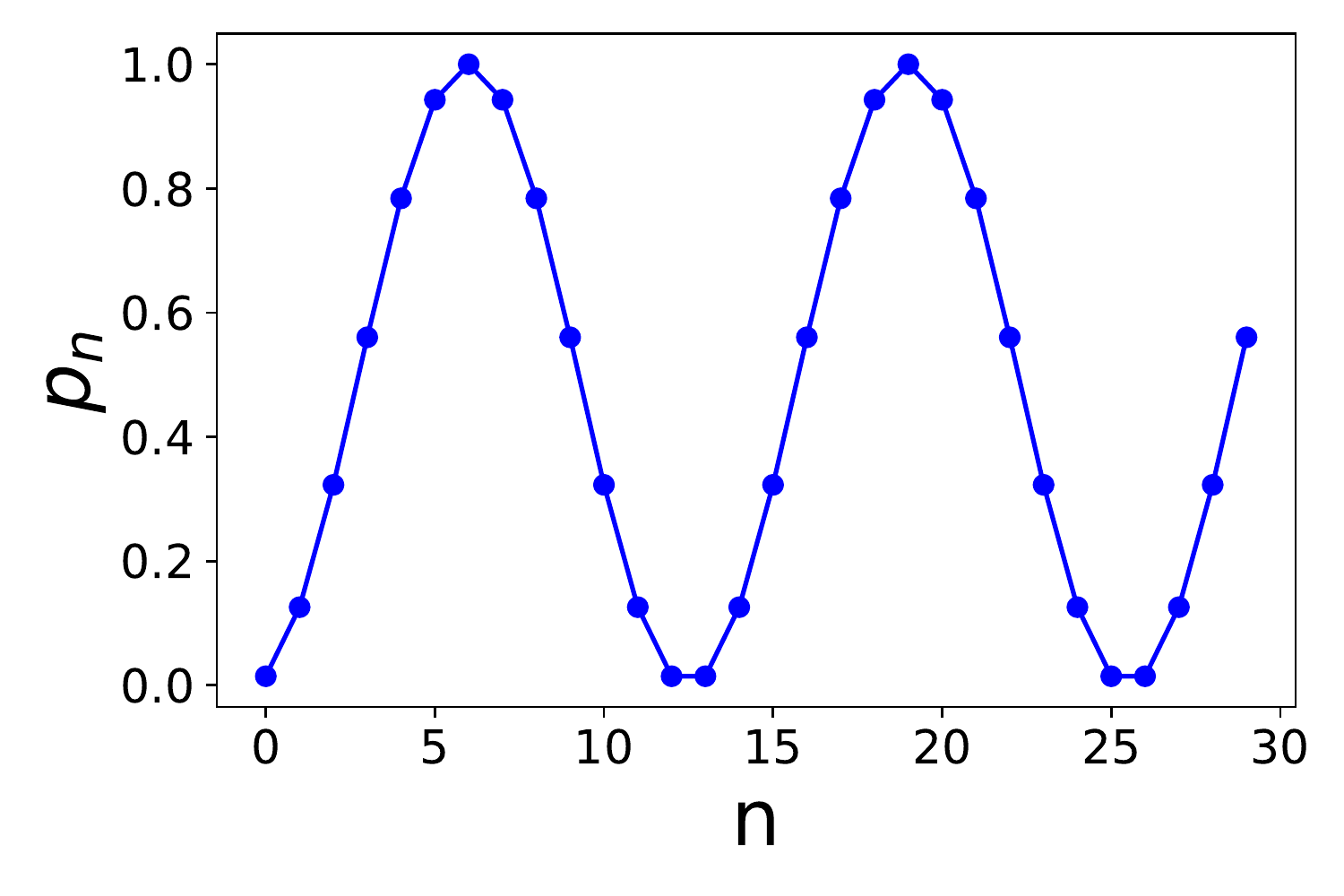}  %width=0.8\linewidth
\par\end{centering}
\caption{Illustration of the evolution of $p_n(\Psi)$ as a function of the number of times $n$ the Grover operator is applied. In this example, the initial state $|\psi \rangle $ is described over 8 qubits. The good state (resp. bad state) is defined as $|\psi_G \rangle = \alpha \sum_{k \in \Omega} |k\rangle$ (resp. $|\psi_G \rangle = \beta \sum_{k \notin \Omega} |k\rangle$), where $\Omega$ is the set of computational states with 4 particles, i.e, $A=4$ and $\alpha, \beta$ are constants that are set so the angle between the initial state $|\psi \rangle$ makes an angle with the bad state of $\pi/26$.}
\label{fig:rotgrover}
\end{figure}

Coming back to the problem of symmetry restoration, the ``good" states correspond to the states respecting a certain symmetry and are associated 
with a projector $\hat P_\alpha$ from which the oracle can be deduced  (see discussion in section \ref{sec:oracle}). One can then use the original Grover 
idea and find the optimal value of $n$ that maximizes $p_n(\Psi)$. As an illustration of this strategy for symmetry restoration, we show in Fig.  \ref{fig:rotgrover} an application to the case of a superfluid system. In this application, a BCS state is first prepared on the QC as described in section \ref{sec:sbansatz}. Then the Grover method is used to restore the particle number symmetry.  

For the case of particle number and when the JWT is used, the oracle is relatively straightforward to construct because the projector is diagonal in the computational basis. For a state on this basis, assuming that we want to project onto a given number of pairs $A$, we have:
\begin{eqnarray}
U_f | k \rangle &=& (-1)^{-\delta_{A n_1(k)}} |k \rangle  ,
\end{eqnarray}     
where $n_1(k)$ is the number of ones in the binary representation of $k$. The matrix $U_f$ is then diagonal with only $1$ and $-1$ in its diagonal in the qubit Hilbert space (see also Fig. \ref{fig:oracle}). While being easy to implement on a classical computer, such diagonal matrix appears to be much more difficult to implement in terms of circuit depth on a quantum computer. To implement the oracle associated with particle number projection, we used the Walsh function basis \cite{Wal23} that was discussed in detail in Ref. \cite{Wel14}. However, such a method is rather costly in terms of the number of operations to perform.  

We see in Fig. \ref{fig:rotgrover} that the probability $p_n$ oscillates back and forth between $0$ and $1$. The optimal value of $n$ leads to a probability close to $1$ within a certain tolerance. One can associate a certain angle $\theta_n$ in the circle displayed in Fig. \ref{fig:grover} and obtain a prescription for the optimal $n$ to be used \cite{Kay11,Lim19}. In practice, it is improbable that this 
optimal value gives exactly $p_n (\Psi) = 1$, and therefore, the final state will have a remaining residual component of the state which breaks the symmetry. To address this limitation, it is possible to use the method proposed by Hoyer \cite{Hoy00} and referred to hereafter as the ``Grover/Hoyer" technique, which is a slightly modified version of the Grover algorithm that will always converge fully to the projected state with $p_n (\Psi) = 1$. In this approach, the general rotation for complete convergence of the state $|\Psi \rangle$ of angle $w=\frac{\pi}{2} - \theta$ is decomposed in $n_G$ arbitrary rotations of angle $\lambda=\frac{w}{n_G}$ with $n_G = \lceil \frac{w}{2\theta} \rceil$ and $\theta$ the angle between $|\Psi \rangle$ and $|\Psi_B\rangle $ in Fig. \ref{fig:grover}. Each singular $\lambda $ rotation makes use of a generalize Grover operator \cite{Hoy00}, which in turn uses a generalized version of the oracle operator of Eq. (\ref{eq:def_oracle}) able to apply an arbitrary phase $e^{i\delta}$ to the states that belong to the "Good space". The Grover/Hoyer method will work similarly as the original ethod of Grover displayed in Fig. \ref{fig:grover}, but will, contrary to the Grover technique, exactly stop at $p_n=1$.

\subsection{Projection by indirect measurement}
\label{sec:indirect}
%\textcolor{ForestGreen}{This should be subsection of section 6.}

We present here several methods with various levels of refinement with a common feature: the restoration of broken symmetries is achieved by measuring one or several ancillary qubits entangled with the system.

\subsubsection{Projection by a single Hadamard test using the oracle}
\label{sec:singleh}

In order to illustrate how projection can be made by indirect measurement, the simplest 
case one can imagine is to perform a Hadamard test using the oracle $U_f$. Starting from a state $\ket{\Psi}$, at the end 
of the Hadamard test circuit and prior to the measurement of the ancillary qubit, the total wave-function is given by:
   \begin{eqnarray}
| \Psi_H \rangle &=& \frac{1}{2} \left\{ | 0 \rangle \otimes \left[ I + U_f \right]| \Psi \rangle 
+ | 1 \rangle \otimes \left[ I - U_f \right]| \Psi \rangle \right\} .  \label{eq:psiH}
 \end{eqnarray}
Therefore, if the initial state decomposes as in Eq. (\ref{eq:psi}) and using the fact that good and bad components are associated to eigenvalues of $U_f$ respectively equal to $-1$ and $1$, we simply deduce that:
\begin{eqnarray}
| \Psi_H \rangle &=& \beta_0 | 0 \rangle \otimes | \Psi_B\rangle  
+ \alpha_0 | 1 \rangle \otimes | \Psi_G \rangle . 
\end{eqnarray} 
We, therefore, see that there is a probability $|\beta_0|^2$ (resp. $|\alpha_0|^2$) to measure $0$ (resp. $1$) in the ancillary qubit after the Hadamard test. Most importantly, if we measure $0$ (resp. $1$), after the measurement the system will be automatically projected onto the bad components $|\Psi_B\rangle$ (resp. the good component $|\Psi_G\rangle$).  
 
Therefore, provided that we have a practical method to implement the oracle, the Hadamard test is a relatively straightforward approach to perform symmetry restoration and, more generally, projection 
at a price to add a single qubit to the quantum register. A second aspect that will be common to all methods presented in this section is that some of the ``events" related to measuring $0$ in the above example are not retained. Therefore, to not waste too many events, it is essential to maximize $|\alpha_0|^2$; this is done in general by an optimization of the $\bm{\theta }$ parameters when preparing the initial state.   

\subsubsection{Projection by quantum phase estimation method}
\label{sec:qpe}

The key to the success of the Hadamard test described previously is that the oracle is a unitary operator with only two eigenvalues.
The quantum phase estimation (QPE) method to perform projection and restore some symmetry can be seen as a direct generalization of the above technique when using symmetry operators having more than two eigenvalues \cite{Lac20}.  

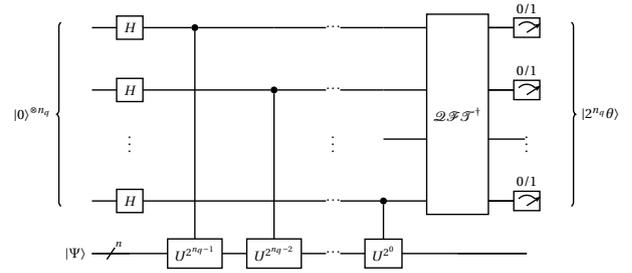
\begin{figure}[htbp] 
\centering
\begin{tikzpicture}
  \node[scale=0.65] {
 \begin{quantikz}
\lstick[wires=4]{$|0\rangle ^{\otimes n_q}$}  & {}             & \gate{H} &  \ctrl{4}             &  \push{}               & \push{} \cdots & \push{}        & \gate[wires=4]{\mathcal{QFT}^{\dagger}} & \meter{0/1} \qw & \rstick[wires=4]{$|2^{n_q}\theta \rangle$}\\
                                              & {}             & \gate{H} &  \push{}              &  \ctrl{3}              & \push{} \cdots & \push{}        & {}                       & \meter{0/1} \qw &\\
                                              & {}                   & \vdots   &  {}                   &      {}                & \vdots         &                &                          & \vdots      \qw &\\
                                              & {}                   & \gate{H} &  \push{}              &  \push{}               & \push{} \cdots & \ctrl{1}       & {}                       & \meter{0/1} \qw &\\
                                              & \lstick{\ket{\Psi}}  & \push{ \qwbundle{n}} &  \gate{U^{2^{n_q-1}}} &  \gate{U^{2^{n_q-2}}}  & \push{} \cdots & \gate{U^{2^0}} & \push{}                  &  \push{}    \qw 
\end{quantikz}
}; 
\end{tikzpicture} \\
    \caption{Schematic view of the QPE method applied to the operator $\hat U$ with $n_q$ ancillary qubits. In the circuit, we took the notation 
    ${\cal QFT}^{-1}$ for the inverse Quantum Fourier transform \cite{Nie02}}.
    \label{fig:quantpe}
\end{figure}

We recall here briefly how the QPE algorithm works. 
The QPE is a method to obtain the eigenvalues of a unitary operator $\hat V$ on a quantum computer \cite{Nie02,Hid19}. 
In the following, the eigenvalues of $V$ are denoted by $e^{i{2\pi} \varphi_k}$, with the condition for all $k$ that is $0 \le \varphi_k < 1$. We show in Fig. \ref{fig:quantpe} the circuit that is used to perform 
the QPE.  We see in this circuit that a set of $n_q$ ancillary qubits is required. 

Let us assume that the initial state decomposes onto eigenstates of $\hat V$, denoted by $ | \phi_k \rangle $ and associated to $\varphi_k$, such that:
\begin{eqnarray}
| \Psi \rangle = \sum_k  \alpha_k | \phi_k \rangle. 
\end{eqnarray} 
After the QPE circuit shown in Fig. \ref{fig:quantpe}  and before the measurement of the $n_q$ ancillary qubits, the total wave-function is given by:
\begin{eqnarray}
| \Psi_{\rm QPE} \rangle &=&  \sum_k \alpha_k | [ \varphi_k 2^{n_q}] \rangle \otimes  | \phi_k \rangle . \label{eq:stateqpe}
\end{eqnarray} 
Here we introduced the notation $[ \varphi_k 2^{n_q}]$ that should be understood as a binary number.  This number is deduced 
from the binary fraction associated to $\varphi_k$ (see Eq. (\ref{eq:binfrac})). Let us consider a specific eigenvalue $\varphi_k = \varphi$ (here we omit 
the index $k$ for simplicity), 
we introduce its binary fraction:\begin{eqnarray}
\varphi &=& 0.\varphi_1 \varphi_2 \cdots \label{eq:binfracphi}
\end{eqnarray}
where the $\varphi_i$ are equal to $0$ or $1$, we have the correspondence $[ \varphi 2^{n_q}] = \varphi_1 \cdots \varphi_{n_q}$. We can 
also define from this binary number an approximation of $\varphi$, denoted by  $\widetilde \varphi (n_q)$ that corresponds to the truncated 
binary fraction:
\begin{eqnarray}
\widetilde \varphi (n_q) &=& 0.\varphi_1 \cdots \varphi_{n_q}. \label{eq:binfractrunc}
\end{eqnarray}
In particular, we have, for a given $n_q$, $| \varphi - \widetilde \varphi (n_q) | < 1/2^{n_q}$.  

The states $| [ \varphi_k 2^{n_q}] \rangle $ correspond to the states of the ancillary register after the computation. We see 
from Eq. (\ref{eq:stateqpe}) that the probability of measuring one of the states is $|\alpha_k|^2$ and that, after the measurement, the system state will be approximately the state $| \phi_k \rangle$. There are several important remarks we can make: (i) if the eigenvalues of $\hat V$ are degenerated, after the measurement, the system state will be a mixture of all degenerated states weighted by their relative contributions in the initial state; (ii) one can get approximations of the eigenvalues $\widetilde \varphi_k (n_q)$ by identifying peaks in the probabilities to measure certain states of the ancillary qubits. The precision of the eigenvalues will depend on $n_q$; (iii) after the measurement, the system state is not perfectly an eigenstate $| \phi_k \rangle$ and can have an admixture of other states. More precisions on this admixture can be found in Ref. \cite{Nie02}.
The only case where a pure eigenstate (or set of degenerated eigenstates) is obtained is when the eigenvalue can exactly be written in terms of a finite binary fraction (\ref{eq:binfracphi}), i.e. where $\varphi_j = 0$ for $j> n_b$, and with the condition $n_q \ge n_j$.

The QPE method was originally designed to obtain eigenvalues or eigenstates that are unknown. Following Ref. \cite{Lac20}, this method can also be used as a projection method to restore symmetries taking advantage of the fact that the eigenvalues of symmetry operators are known and can be associated in general to a set of integer numbers. This is the case for all operators discussed in section \ref {sec:symcons}. 

Let us consider one of these symmetry operators $\hat S$. This operator has a finite discrete set of eigenvalues written in ascending order as $\{\lambda_0 \le  \cdots \le  \lambda_M\}$.  It is assumed that one can connect to these eigenvalues a set of integers $\{m_0 \le  \cdots \le  m_M\}$ through a linear 
relation $\lambda_k = a m_k$, where $a$ is a constant. We then introduce the operator $\hat V$ to be used in the QPE shown in Fig. \ref{fig:quantpe}. 
A possible choice for this operator is \cite{Lac20,Siw21}: 
\begin{eqnarray}
\hat V &=& \displaystyle  \exp\left\{ 2\pi i \left[ \frac{\hat S - \lambda_0}{  a 2^{n_0} } \right] \right\} .  \label{eq:vs}
\end{eqnarray} 
We can then associate the phase $\varphi_k = (m_k - m_0)/2^{n_0}$ to each eigenvalue of $\hat V$. All phases should verify 
$0 \le \varphi_k  < 1$. This gives the condition:
\begin{eqnarray}
\ln(m_k - m_0)/\ln 2 < n_0.  \label{eq:ineq}
\end{eqnarray} 
With this condition, all phases $\varphi_k$ are automatically finite binary fractions as written in Eq. (\ref{eq:binfractrunc}) provided that $n_q = n_0$. 
In practice, it is convenient to minimize the number of ancillary qubits and take the lowest value $n_0$ for which the inequality 
(\ref{eq:ineq}) is verified. 

Applying the QPE approach with the appropriate number of qubits together with the operator (\ref{eq:vs}), each eigenvalue $\lambda_k$ 
becomes automatically associated to one of the states $| [ \varphi_k 2^{n_q}] \rangle$ in Eq. (\ref{eq:stateqpe}), while the state 
$| \phi_k \rangle $ is the projected state on the part of the Hilbert space associated with this $\lambda_k$. This implies that it corresponds to the symmetry restored state with a specific value of the symmetry operator eigenstate. 
Therefore, measuring the ancillary register with the outcome $[ \varphi_k 2^{n_q}]$ acts as a symmetry restoration. After the measurement of one of the channels, the system state is automatically projected onto one of the subspaces and respects the symmetry associated with $\hat S$. This method was called discrete spectra assisted (DSA) approach in Ref. \cite{Lac20}.
It achieves a slightly different goal than the method presented in section \ref{sec:singleh}. Since, here, all channels associated a priori to the complete set of projectors $\hat P_\alpha$ are simultaneously accessible. Indeed, in the QPE-based projection approach, each measurement can correspond to one of the $\hat P_\alpha$ that changes from one event to the others. This parallelism was used in Refs. \cite{Lac20,Siw21} to get all projections with the same circuit either for the projection on particle number or total spin. 

\subsubsection{Projection by a set of independent Hadamard tests: the Iterative QPE (IQPE)-like method }
\label{sec:iqpe}

\begin{figure}[htbp] 
\begin{center}
\includegraphics[width=0.7\linewidth]{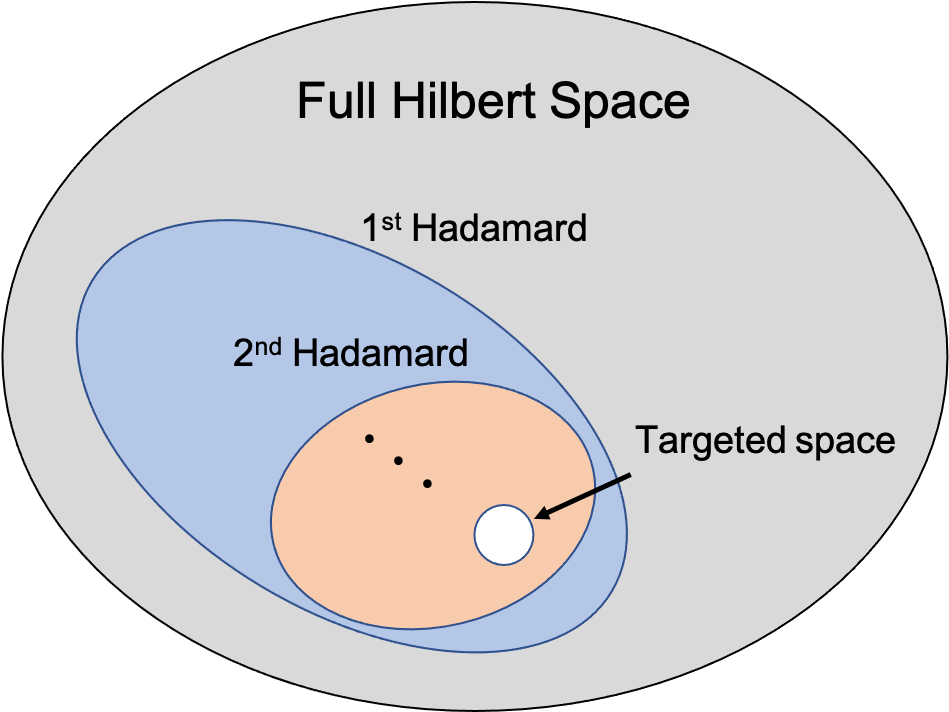} %width=\columnwidth
\end{center}
    \caption{Schematic representation of the iterative method based on a set of successive IQPE circuits that leads to a projection on a targeted sector of the entire Hilbert space. In this figure,  the collection of IQPE circuits progressively projects the initial state on a more and more reduced part of the total Hilbert space until this space reaches the desired subspace.}
    \label{fig:nhadamard}
\end{figure}

We conclude the set of methods based on the indirect measurement by a technique that could be regarded as an intermediate approach between the QPE approach and the approach based on the combination of a single Hadamard and an oracle.

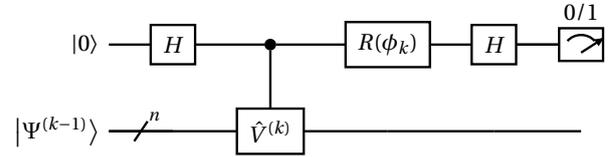
\begin{figure}[htbp]  
\centering
\begin{tikzpicture}
  \node[scale=1.1] {
\begin{quantikz}
\lstick{\ket{0}}            & \gate{H}               & \ctrl{1}                      & \gate{R(\phi_k)}  & \gate{H}  & \meter{0/1} \qw  \\ %next line
\lstick{\ket{\Psi^{(k-1)}}} & \push{ \qwbundle{n}}   & \gate[wires=1]{ \hat V^{(k)}} & \push{} & \push{} &  {} \qw
 \end{quantikz} 
}; 
\end{tikzpicture} \\
    \caption{Illustration of the circuit called ``IQPE-like circuit" in the main text. This circuit corresponds to a modified Hadamard test where a phase is added $e^{i\phi_k}$ is added to the 
    ancillary qubit component $|1\rangle$ through the phase operator $R(\phi_k)$ (see text). }
    \label{fig:circkitaev}
\end{figure} 
In this approach, we use $n_{\rm H}$ successive circuits, each using one ancillary qubit. The method is similar to the one proposed by Kitaev in Ref. \cite{Kit95} or to the Iterative QPE(IQPE) Ref. \cite{Dob07}. We will call it the IQPE-like projection method below because of the resemblance of the circuit used with the one in the IQPE algorithm. Each of these circuits is essentially a Hadamard test with an additional phase gate of parameter $\varphi$ before the second Hadamard gate of the ancillary qubit. The corresponding circuit is shown in Fig. \ref{fig:circkitaev} and will be referred to as the IQPE-like circuit below. The $n_{\rm H}$ successive IQPE-like circuits are associated with a set of operators $\hat V^{(k)}$ with $k=1,\cdots,n_H$ that allow arriving progressively to the targeted subspace. Fig.  \ref{fig:nhadamard} shows a schematic view of the technique. At step $k$, i.e., after applying $k$ circuits, we write the system state similarly to Eq. \eqref{eq:psi}  with: 
\begin{eqnarray}
| \Psi^{(k)} \rangle &=& \alpha_k | \Phi^{(k)}_{\rm Keep} \rangle  + \beta_k   | \Phi^{(k)}_{\rm Trash} \rangle \label{eq:keeportrash}
\end{eqnarray}      
where $\ket{ \Phi^{(k)}_{\rm Keep}}$ is a part of the wave-function that we want after the next Hadamard test while $| \Phi^{(k)}_{\rm Trash} \rangle$ will be removed at the $(k+1)^{th}$ step. This iterative method is stopped when the resulting state identifies with the good component $| \Psi_G \rangle$. A condition for the success of the method is that the targeted state is always included in the component retained at each step (see Fig. \ref{fig:nhadamard}). Noteworthy, there is no reason a priori that the set of operators $\{ \hat V^{(k)} \}$ is unique, and hence, a proper choice of the operators can be used to reduce the number of Hadamard tests or the circuit depth or both. We note also that, for the particle number case, the projector written as in Eq. \eqref{eq:prodN} gives some guidance on how such iterative procedure can be operated (see also discussion in \cite{Ata73,Kha21}).

Let us consider the particle number case and assume that we use a set of circuits with one of the operators $\hat V^{(k)} = e^{i\phi_k \hat N}$ and additional phase $\varphi = -\phi_k A$ where $A$ is the targeted number of particles and $\phi_k = \pi/2^{k}$ with $k=0, 1, \cdots, n_H$. The eigenstates of this operator are those of the natural basis. The associated eigenvalues are shifted due to the additional phase and can be considered effectively equal to $e^{i \phi_k (n_1 - A) }$, where again $n_1$ denotes the number of $1$ in the binary representation of a given NB state. Let us consider the case $k=0$, for which the eigenvalues become $e^{i \pi (n_1 -A)}$ and equals $1$ (resp. -1) if $(n_1 -A)$ is even (resp. odd).  Therefore, if we use a IQPE-like circuit with the $k=0$ case, according to Eq. (\ref{eq:psiH}), if we measure $0$ in the ancillary qubit, all odd values of  $(n_1 -A)$ are removed and only even values will be kept in the system wave-function. If we use a second circuit with $k=1$, eigenvalues become $e^{i \pi (n_1 -A) /2}$ and all $n_1$ with $(n_1 -A) /2$ being odd are removed if we measure $0$ in the ancillary qubit of the second circuit. Then the procedure can be iterated such that after measuring $0$ with the $k^{th}$ IQPE-like circuit, we remove all eigenstates where  $(n_1 -A) /2^k $ is odd. The process stops when all values of $(n_1-A)$ are removed except the one with $n_1=A$, i.e. when we have kept only the component of the initial state with exactly $A$ particles.

The present procedure can be extended to any symmetry restoration provided that all eigenvalues of the symmetry operator $\hat S$ can be connected to a set of integers $\{m_0 \le  \cdots \le  m_M\}$ as in section \ref{sec:qpe} (with $\lambda_k = a m_k$). Assuming that we want to project the initial wave-function on the subspace associated with the eigenvalue $\lambda_\alpha$. This could be done using a set of operators 
\begin{eqnarray}
\hat V_l &=& \displaystyle  \exp\left\{ i \frac{\pi}{a2^l}  \left[\hat S - \lambda_\alpha \right] \right\} .  \label{eq:vlhad}
\end{eqnarray} 
with $l=0, 1, \cdots , n_H - 1$, where the value $n_H$ is finite and will depend on ${\rm max}(m_\alpha - m_0, m_M - m_\alpha )$. After the set of IQPE-like circuits, only events measuring a sequence of $0$s in the iterative measurement are retained. 
\begin{figure}[htbp] 
\begin{center}
\includegraphics[width=\columnwidth]{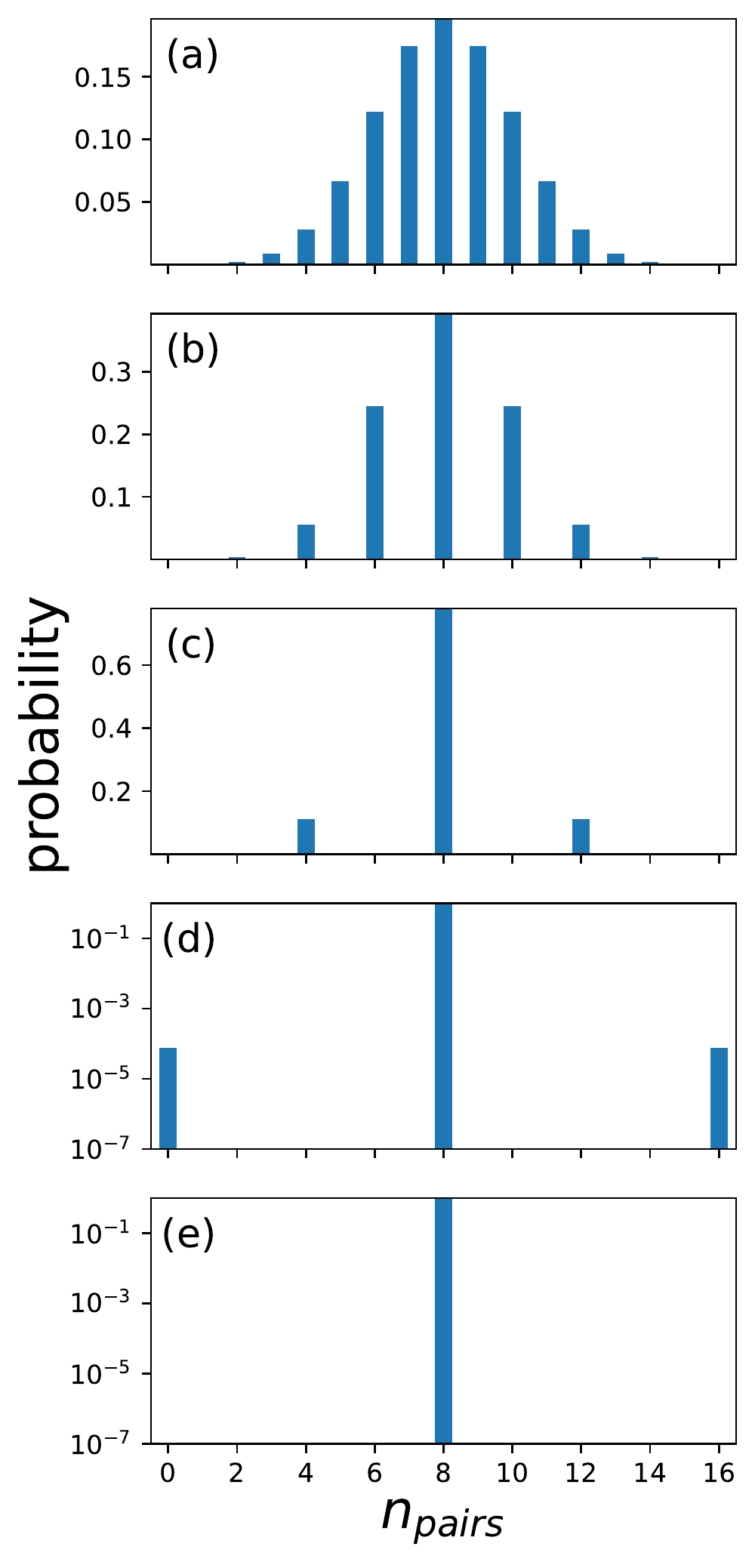}  %width=0.8\linewidth
\end{center} 
    \caption{Illustration of the iterative method based on a set of IQPE-like circuits to perform projection with an initial state constructed as an equiprobable sum of the NB states. 16 qubits are used to describe the system. The different panels show the amplitudes of the 
    state associated to various numbers of pairs $n_{\rm pairs}$ 
    at intermediate stages of the IQPE-like method. The method is applied for the projection to 
    $n_{\rm pairs}=8$. Panel (a): amplitude of the initial state. Panel (b): amplitudes after the application of the first circuit. After the measurement of the first ancillary qubit, all amplitudes with odd values of $n_{\rm pairs}$ are zero. Panels (c), (d) and (e) correspond to the amplitudes after using the first two, three and four 
    IQPE-like circuits respectively. We see that all components disappear progressively except the targeted one. Note that the last two panels ((d) and (e)) are plotted on a logarithmic scale so one can see the disappearance of the final two components that are of the order of $10^{-5}$.}
    \label{fig:sechadamard} 
\end{figure}
An application connected to the present approach can be found in Ref. \cite{McA19}. The current approach has several attractive aspects compared to the two previous techniques.
Firstly, it requires only one qubit at a time and does not necessitate performing the inverse Fourier transform, as is the case for the previous QPE method. The iterative process can easily accommodate the simultaneous projection of several symmetries. Finally, it is worth noting that in the many-body problem, coherent states like BCS, after optimization imposing 
a certain value of particle number $A$ in average, often have a Gaussian probability decaying fast around $A$ for the amplitudes related to states with particle number different from $A$ \cite{Bla86}. Because of this exponential decay, it is anticipated that very few IQPE-like circuits will be necessary to remove the unwanted components 
of the wave-function to an excellent approximation, and, ultimately, the full set of IQPE-like circuits will not be required. 

We illustrate in Fig. \ref{fig:sechadamard} the effect of the iterative projection starting from a many-body state formed by an equiprobable population of the NB states.     
Note that, compared to the method based on a single Hadamard test combined with the oracle, the present approach uses relatively simple 
unitary operators that are easier to implement than the oracle. Altogether, we have observed in practice that the last method we have presented seems to be the best compromise between the number of required ancillary qubits and the complexity of the circuit required to perform the symmetry restoration.  

We finally mention that another approach based on an iterative set of Hadamard tests was proposed in Ref. \cite{Siw21}. The latter method differs, however, from the one we present here and is more subtle because (i) it is based on a property of total spin operators that allows defining a new set of operators where the number of permutations scales linearly with the number of particles while it was quadratic for the total spin and (ii) the operator to be used at the step $(k+1)$ depends on the results of the Hadamard test at step $(k)$.  

\section{Comparison between the different symmetry restoration methods for the particle number symmetry}
\label{sec:comparison}

We have presented several methods that might be used to access some properties of symmetry-restored states or the projected states with proper symmetries. These methods are sometimes quite different in terms of the concepts used, which translate into differences in the quantum resources required to implement them, which is crucial for future implementations. 
In this section, we make a discussion specifically focused on comparing the resources. 
We use the case of particle number symmetry as an illustration. Multiple aspects should be considered when comparing different quantum algorithms for symmetry restoration. We give below a systematic discussion of these aspects, together with a summary of some of these aspects in table \ref{tab:comparison_projections}.
%Each component of the different algorithms is discussed one after the other below:

\begin{itemize}
%\item[(a)] {\bf Goal of the method:} First, while the different methods presented here target the symmetry-restored state that could be used for further post-processing, the method based on Eq. (\ref{eq:oexp}) is less ambitious. It gives access to expectation values of observable without constructing the SR state itself. This fact should be kept in mind for the comparison below.
\item[(a)] {\bf Size of the ancillary register:} Differences exist between the different methods regarding the number of ancillary qubits that should be added to the system register. The least demanding technique is the Grover/Hoyer method, which does not require any additional qubit. On the other hand,  the two most demanding are the QPE method and the LCU projection technique, which both require the simultaneous use of several qubits proportional to $\log_2 n_q$ where $n_q$ is the number of qubits for the system. The exact numbers 
of ancillary qubits in these two cases are given in table \ref{tab:comparison_projections}. 
%\textcolor{red}{(Andres, I do not understand what do you want to say here) The IQPE-like method presented in section \ref{sec:iqpe} slightly simplified the QPE approach.} 
Similarly to the QPE, the IQPE-like approach presented in section \ref{sec:iqpe} uses a set of ancillary qubits, but each can be used one after the other. Then, assuming that reusable qubits are available, only one extra qubit that is used several times is needed.  
The direct use of Eq. (\ref{eq:oexp}) also requires one extra qubit at a time to obtain the set of expectation values entering in these equations.

\item[(b)] {\bf Preparation and post-treatment of the ancillary qubits:} For methods that require ancillary qubits, i.e., all but the Grover-based method, the operations involving the extra qubits can be separated into two classes. One class of operations are made to transfer information from the system to the ancillary qubits (controlled operations).The other class corresponds to the operations 
required before or after the controlled operations to prepare or post-process the ancillary qubits before measurements. We focus here on the latter sets of operations. The method of Eq. (\ref{eq:oexp}), together with the IQPE-like and the Grover+Hadamard methods, are based on Hadamard or modified Hadamard tests. The QPE and LCU-based methods are much more costly regarding operations directly applied to the ancillary qubits. For instance, from Ref. \cite{Nie02}, the $QFT^{\dagger}$ used in the QPE (see Fig. \ref{fig:quantpe}) can be implemented using $\mathcal{O}\left(n^2\right)$ Hadamard gates and controlled phase gates with $n$ as the number of qubits in the register where the gate is being implemented. 
%% I removed this discussion that is not absolutely necessary in view of our level of discussion. 
%\textcolor{ForestGreen}{There exists some approximations which achieve an scaling of %$\mathcal{O}\left(n \log \left( n\right)\right)$ \cite{Hal00,Nam20} [This line is not clear. Under approximations which scaling is logarithmic, Hadamard gates or controlled phase gates?]}\textcolor{brown}{I had an error in the scaling of the QFT. Now that it is corrected, I think the phrase makes sense. The answer to the question is both, both Hadamard gates and control phase gates achieve and scaling of $\mathcal{O}\left(n log_2 n\right)$}. 
In the LCU-based method (Fig. \ref{fig:projLCU}), the ancillary register is prepared with the operation $B$ and post-processed with the operation $B^\dagger$. Because of this, the operations on the ancillary register for the LCU are extremely costly regarding quantum resources. 

\item[(c)] {\bf Controlled operations between the ancillary and qubit registers:} 
%The addition of ancillary qubits is always associated with controlled operations to transfer information from the system to the ancillary register. 
In the QPE, the IQPE-like or LCU-based, for the particle number case, the controlled operations are made on the system using operators of the form $e^{i\phi_k \hat{N}}$ appearing in Eq. \eqref{eq:pngen4}. These operators can be decomposed as the tensor product of phase gates \cite{Rui21}, i.e.:
\begin{eqnarray}
e^{i\phi_{k}\hat{N}}=\bigotimes_{m=0}^{n_q -1}
\begin{pmatrix}1 & 0\\
0 & e^{i\phi_{k}}
\end{pmatrix}_m . 
\end{eqnarray}
This decomposition implies that each operator requires $n_q$ controlled phase gates. For the QPE, IQPE-like, and LCU, these operators are used $n_{QPE}$, $n_{IQPE}$, and $n_{LCU}$ times (see table \ref{tab:comparison_projections} for further the definition of this numbers), respectively. 
In the QPE and IQPE methods, each phase gate is controlled by one ancilla qubit, different from the LCU method, where each phase gate is controlled simultaneously by $n_{LCU}$ ancilla qubits. 
%\textcolor{red}{(Andres: I don't think we should say this phrase. We never discussed how to decompose the projector in Pauli strings) The direct application of Eq. (\ref{eq:oexp}) is much less demanding in terms of number of controlled operations since each calculation has only one controlled sequence with one of the Pauli string $\Lambda$ appearing in the LCU decomposition of the projector and that enters in given in Eq. (\ref{eq:oexp})}. Last, for the Grover+Hadamard case, \textcolor{red}{(Andres: I do not understand this phrase) the operator with which the controlled operation is directly the oracle that requires a large set of operations (see item (e) below).}
\item[(d)] {\bf Processing the system state only:} Most methods above require operations made on the system through controlling the ancillary qubits; this is the case for the QPE, IQPE, and LCU-based. 
The Grover/Hoyer method is at variance with these methods since, in its simplest form, it does not require any ancillary qubit. 
In the Oracle + Hadamard method, we also see that only one qubit is requited. 
%\textcolor{red}{(Andres: I do not understand what do you want to say here. In fact, I'm not sure what is the point of the whole (d) section.). The counterpart of this interesting feature for oracle-based 
%approaches the necessity to directly manipulate the system state by applying, for instance, the oracle itself.}
However, the drawback is the necessity to explicitly construct the Oracle that is costly in operations (see discussion in section \ref{sec:amplitude}).
%The oracle is diagonal in the qubit computational basis for the particle number case. 
%Refs. \cite{Bul04,Wel14} show that the implementation of an arbitrary diagonal operator requires $2^{n_q+1}-3$ one and two-qubit gates. This scaling is the main restriction that prevents the use of the \textcolor{red}{(Andres: Again, are you referring to the Grover/Hoyer method?) Oracle-based method today.}
\item[(e)] {\bf Measurements and wasted events:} Most methods are based on the "rejection until success" of events. For example, the LCU-based approach retains only one type of event, those where only zeros are measured in all ancillary qubits. 
This feature implies that some events are wasted and should not be retained. In general, the number of wasted events is directly proportional to $(1-p_G)$ where $\sqrt{p_G}$ is the amplitude of the initial state on the "Good state", i.e., the amplitude of the component with the proper symmetry. The table \ref{tab:comparison_projections} shows that the QPE-, LCU-, and Oracle-based methods are essentially equivalent in terms of the number of retained events. Only in the Grover/Hoyer method, the projected 
state is obtained with 100 \% probability provided that $p_G \neq 0$.  
%\textcolor{ForestGreen}{[How does this factor affect the performance of an algorithm? If it does not, maybe we do not need this section.]}\textcolor{brown}{We can understand the importance of mentioning that these three algorithms are probabilistic (they will work with a probability $\sim p_G$) instead of all the time by comparing with the Grover/Hoyer case. Assuming $p_G\neq 0$, the Grover/Hoyer algorithm will produce a projected state after applying the gates. In contrast, the three methods mentioned, as they included probabilistic measurements, will not give the projected state each time the algorithm is performed. I do not know if we should include some discussion in the text given that what I said is already mentioned; see section \ref{sec:amplitude}.}
%\textcolor{red}{(Andres; I do not understand this. All measurements in the IQPE-like method should be 0 to obtain full projection. Why would we stop in an intermediary state? Plus, I'm not sure if we stop when state $|1\rangle$ is measured; we would have the filtering of the states up to that point.) Still, the IQPE-like approach can be stopped at an intermediate stage as soon as one qubit is measured in the state "1". }
\end{itemize}

%\textcolor{brown}{To make a comparison, we take the case of the number of particles symmetry. This symmetry is a straightforward case because the projector and the oracle are diagonal in the computational basis. Next, we will discuss the resources necessary to implement the different quantum subroutines that the algorithms employ. The relation between these subroutines and the projective methods is presented in Table \ref{tab:comparison_projections}.}

We see above that some methods are more advantageous than others depending on which aspects in (a-e) are considered. The exponential scaling of the Quantum State Preparation (QSP) algorithm, used in the LCU-based and Grover/Hoyer methods, is by far the most demanding in terms of quantum resources compared to the other quantum subroutines. This scaling means that the Grover/Hoyer and LCU-based methods can be directly assigned as the most costly approaches. The cost of implementing the oracle follows QSP as the most expensive subroutine; its implementation qualifies the Oracle + Hadamard as the third most costly algorithm for projection. Finally, implementing the inverse QFT and the multiple uses of ancillary qubits rates the QPE algorithm as the fourth most expensive approach. Two approaches then appear as the best compromise today for symmetry restoration, depending on whether or not the projected state is required. In the former case, it is the IQPE-like method. In the NISQ context, the most suitable technique is reconstructing expectation values of observables by Eq. (\ref{eq:oexp}).

\begin{table*}[t]
  \centering
  \resizebox{\textwidth}{!}{%
    \begin{tabular}{ |c|c|c|c|c| } 
    \hline
     \backslashbox{Method}{Characteristic} & $\#$ ancilla & $\#$ measurements & Gate ressources & \makecell{Access to\\ the projected state} 
     %& \makecell{Ranking cost\\ of implementation}
     \\
     \hline
     \thead{Post-processing Eq. (\ref{eq:oexp})} & \makecell{1} & - & \makecell{$\Lambda$ Hadamard tests, each:\\
                                                                                - 2 Hadamard gates.\\
                                                                                - 1 control operator.\\
                                                                                - 1 controlled evolution operator.}&No%&-
                                                                                \\
     
     \hline
     \thead{QPE (section \ref{sec:qpe})} & $ n_{QPE} = \left\lfloor log_{2} \left( n_q\right) \right\rfloor$ & $\sim p_G$ & \makecell{ $\Lambda$ Controlled-evolution gates.\\
                                                                                    $n_{QPE}$ Hadamard gates.\\
                                                                                    An inverse QFT over $n_{QPE}$ qubits.}&Yes%&3
                                                                                    \\

     \hline
     \thead{LCU (section \ref{sec:lcu})} & $ n_{LCU} = \left\lceil log_{2} \left( \Lambda + 1\right)\right\rceil$ & $\sim p_G$ & \makecell{ $\Lambda$ Controlled-evolution gates.\\
                                                                                    2 QSP circuit over $n_{LCU}$ qubits.}&Yes%&4
                                                                                    \\
    \hline
    
    \thead{Grover/Hoyer (section \ref{sec:amplitude})} & 0 & - & \makecell{2 generalized oracles \\
                                          $n_{G}$ generalized Grover operator, each:\\
                                          - 1 generalized oracle. \\
                                          - 2 QSP over $n_q$ qubits.\\
                                          - 1 phase gate controlled by $n_q - 1$ qubits.\\
                                          }&Yes%&5
                                          \\
     \hline
     \thead{IQPE-like (section \ref{sec:iqpe})}& 1 & $\sim p_G$ & \makecell{At most $n_{IQPE}$ IQPE-like circuits, each:\\
                                                - 2 Hadamard gates.\\
                                                - 1 Controlled evolution gates.\\
                                                - 1 Phase gate.}&Yes%&1
                                                \\
     \hline
     \thead{Oracle+Hadamard (section \ref{sec:singleh})}& 1 & $\sim p_G $ & \makecell{1 controlled oracle\\
                                                          2 Hadamard gates}&Yes%&2
                                                          \\
     \hline
    
    \end{tabular}
    }
\caption{ Some elements of comparison between the different projection methods presented in this article. From the left to the right column are shown the number of ancillary qubits, the probability of retaining an event in the measurement, a summarized discussion of the quantum subroutines required to implement the method, and if the approach gives access to the projected state or not. In this table, $\Lambda$ is the number of terms in the LCU form of the projector provided by Eq. (\ref{eq:LCU}). 
For the number of particles $\Lambda = n_q + 1$ with $n_q$, the number of qubits in the principal register. $p_G=\langle \Psi_G | \Psi \rangle$ is the fraction/amplitude of the desired symmetry (Good component). In the Grover/Hoyer method, we have $n_G=\lceil \frac{w}{2\theta} \rceil$ with $w=\frac{\pi}{2} - \theta$ and $\sin^{2}\left(\theta \right) = p_G$. Details of the generalized oracle mentioned in the Grover/Hoyer method are discussed in Ref. \cite{Hoy00}. 
In the IQPE-like method, we have $n_{IQPE} = \lfloor \log_{2} \left( n_q \right) \rfloor + 1$.    
We also mention that it is possible to implement a less costly version of the LCU in which one state preparation subroutine is changed by a set of $n_{LCU}$ Hadamard gates \cite{Wei20}.}
\label{tab:comparison_projections}
\end{table*}

\section{Conclusion} 

We give an overview of different aspects of treating many-body problems and their symmetries on quantum computers. We start here from a rather didactic overview of different facets of symmetries in a quantum problem that can be useful for experts or non-expert in this field.
This overview includes a discussion of operators associated with the most common symmetries and the corresponding projectors that allow selecting only the states having these symmetries. Specific attention is given to the possibility of considering symmetry preserving states because, as in the case of classical computers, taking advantage of the symmetry of the problem is a way to focus automatically on the relevant parts of the total Hilbert space. In the case of a quantum computer, this allows reducing the number of qubits to describe a system significantly. For some specific correlations sometimes occurring in interacting particles, for instance, the superfluidity effect, it might be potent to consider symmetry breaking state. Breaking symmetry is relatively straightforward on a quantum computer. Still, the price to pay is the 
necessity to restore the symmetry a posteriori when a precise description of the correlated system is required. In the present 
work, we give a panel of possible methods that aim at symmetry restorations. 

In particular, we complete the previously proposed strategies, based on adding ancillary qubits together with indirect measurements, with two new approaches to perform symmetry restoration. The first one is based on the direct application of the projector on the good symmetry space. The second method, which does not require additional qubits, is inspired by the quantum search
algorithm and uses the oracle concept. In the latter, the wave function is gradually purified to remove the unwanted symmetry-breaking components of a given initial state.

These methods differ in the quantum resources needed to implement them (number of qubits and circuit depths), and we expect that their use for future applications will be facilitated by the progress made by future quantum platforms. We also make here a comparison of quantum resources between the different symmetry restoration methods.

We finally note that a set of alternative algorithms have been proposed in Ref. \cite{Lab21} to test symmetries either at the level of the state vector in connection with entanglement or directly at the level of Hamiltonian in \cite{Lab22}.  

\appendix

\section{Antisymmetrization in fermionic systems treated in quantum computers}
\label{sec:fermion}

One of our target applications of symmetry problem are many-body problems formed of particles 
that interacts with each other Specifically, we are more interested in fermionic systems 
that are anti-symmetric with respect to the exchange of two particles. We give a short discussion here 
on how Fermi systems are treated in quantum computers. 
A standard way to treat this problem is to map the fermionic Fock space 
into the qubit Hilbert space. Let us consider a system describing a set of fermions. Using second quantization formalism, any operator 
can be written in terms of creation/annihilation operators $(a^\dagger_j, a_j)$ associated to a complete set of single-particle states $| j \rangle$ and defined with respect to a vacuum state $| - \rangle$. For fermionic systems, we have $\{ a_j , a^\dagger_k \} = a_j  a^\dagger_k + a^\dagger_k a_j = \delta_{kj}$ and for any state: 
\begin{eqnarray}
a^\dagger_j | - \rangle &=& | j \rangle ~~{\rm and} ~~ a_j |- \rangle =0. 
\end{eqnarray} 
The mapping from fermions to qubits is often made in such a way that the $s_j = 1$ (resp. $0$) is interpreted as the occupation (vacancy) of the qubit $| j \rangle$.  Therefore, the NB state $\bigotimes_j | 0_j\rangle$ identifies the Fock vacuum. For the sake of compactness, we will sometimes use $|-\rangle$ also for this specific state of the NB. With this convention, we find that the fermionic creation operator is transformed to a qubit operator as: 
\begin{eqnarray}
Q^+_j &=& \frac{1}{2} \left( X_j - i Y_j \right) =  \left[ \begin{array}{rr} 0& 0\\ 1&0 \end{array}\right]  ,
%a^\dagger_j &\longrightarrow &  Q^+_j \otimes Z^{<}_{j-1} , \label{eq:jwt}
\end{eqnarray}
that fulfills the wanted relationships 
\begin{eqnarray}
Q^+_j | 0_j \rangle=| 1_j \rangle, ~Q_j  | 0_j \rangle = 0,
\end{eqnarray} 
together with $\{Q_j, Q^+_j \}=1$. 
If more than one qubit is considered, we have to ensure that the anticommutation relation between fermions holds. However, for $j \neq k$, the operator defined above commutes, i.e. $[Q_j , Q^+_k] = 0$. This
difficulty can be overcome using the Jordan-Wigner transformation (JWT) that corresponds to the mapping \cite{Jor28,Lie61,Som02,Fan19}: 
\begin{eqnarray}
a^\dagger_j &\longrightarrow &  Q^+_j \otimes Z^{<}_{j-1} , \label{eq:jwt}
\end{eqnarray}
with the definition $Z^{<}_{j-1} =  \bigotimes_{k=1}^{j-1} (-Z_k)$. This mapping solves the anticommutation problem which arises when mapping the indistinguishable fermions on the distinguishable qubits. We mention that the mapping from fermions to qubits is not unique and that alternative mappings have been proposed \cite{Bra02,See12,Bau20}.

\section{Proof of the equivalence between (\ref{eq:prodN}) and (\ref{eq:pngen4})}
\label{app:proofsprojN}

We start from Eq. (\ref{eq:prodN}). Let us assume that we take $M$ as being a power of $M=2^n$ such that 
$2^n \ge {\rm max} (k-N)$, where $k=0, \ldots, \Omega$ and $\Omega$ stands for the maximal value of $k$ (maximal number of particles). 
Let us write $k$ as a binary fraction. For all $k \le M-1$, we can write 
(using the convention of Ref. \cite{Nie02} page 218):
\begin{eqnarray}
\frac{k}{2^n} &=& 0.k_1 \cdots k_n = \frac{k_1}{2} + \cdots \frac{k_n}{2^{n}}, \label{eq:binfrac}
\end{eqnarray} 
where the $k_i$ are equal to zero or one. Then, we use the following chain of identities (here $\hat j$ is replacing $\hat N - N$ for compactness): \begin{eqnarray}
\sum_{k=0}^{M-1} e^{2 \pi i k \hat{j} /2^{n}} &=& \sum_{k=0}^{M-1} e^{2 \pi i \hat j 0.k_1\cdots k_n } 
= \sum_{k=0}^{M-1} e^{2 \pi i \hat j \sum_{l=1}^{n}  k_l 2^{-l} } ,\nonumber \\
&=&  \sum_{k=0}^{M-1} \prod_{l=1}^{n} e^{2 \pi i \hat j k_l/2^{l}} . \nonumber 
\end{eqnarray} 
Since $k$ takes all values between $0$ and $M-1$, we can write:
\begin{eqnarray}
\sum_{k=0}^{M-1} e^{2 \pi i k \hat  j /2^{n}} &=& \sum_{k_1 = 0}^{1} \cdots \sum_{k_n=0}^{1} \prod_{l=1}^{n} e^{2 \pi i \hat j k_l/2^{l}}, \nonumber \\
 &=& \prod_{l=1}^{n} \left( I + e^{\frac{\pi}{2^{l-1}} i \hat j } \right)  \equiv  \prod_{l=0}^{n-1} \left( I + e^{ i \phi_l \hat j } \right) , \nonumber 
\end{eqnarray}
with $\phi_l = \pi / 2^{l}$ and $l=0, \cdots , n-1$. If we account for the $1/M$ factor, we have deduced indeed that we have a formula 
 \begin{eqnarray}
 P_N &=&  \prod_{l=0}^{l_{\rm max}} \frac{1}{2} \left( 1 + e^{i\phi_l (\hat N - N)}\right), \label{eq:product}
 \end{eqnarray}
 proving the equivalence with Eq. (\ref{eq:pngen4}).

\begin{acknowledgements}
This project has received financial support from the CNRS through
the 80Prime program and is part of the QC2I-IN2P3 project. This work was supported in part by the U.S. Department
of Energy, Office of Science, Office of High Energy Physics,
under Award No. DE-SC0019465. We acknowledge
the use of IBM Q cloud as well as use of the Qiskit software package
\cite{Abr19} for performing the quantum simulations.
\end{acknowledgements}


\begin{thebibliography}{99}


\bibitem{Gro96} D. J. Gross, {\it The role of symmetry in fundamental physics},  Proc. Natl. Acad. Sci. USA, {\it 14256} (1996). 


\bibitem{Wei95a} S. Weinberg, {\it The quantum theory of fields (Vol. I)} (Cambridge university press, 1995).

\bibitem{Wei95b} S. Weinberg, {\it The quantum theory of fields (Vol. II)} (Cambridge university press, 1995). 


%%% SPINS 

\bibitem{Mes62} A. Messiah, {\it Quantum mechanics, vol. II}. (North-Holland, Amsterdam, 1962).

\bibitem{Kap75} I.G. Kaplan, {\it Symmetry of Many-Electron Systems}, Academic Press New York, (1975). 

\bibitem{Ham12} M. Hamermesh, {\it Group Theory and Its Applications to Physical Problems}, Courier Corporation, 2012.

%%%%%% General QC


\bibitem{Fan19} G. Fano and S. M. Blinder, {\it Quantum chemistry on a quantum computer}, in Mathematical Physics in Theoretical Chemistry (Elsevier, Amsterdam, 2019), pp. 377–400.


\bibitem{Cao19} Y. Cao et al., {\it Quantum chemistry in the age of quantum computing}, Chem. Rev. {\bf 119}, 10856 (2019).


\bibitem{McA20} S. McArdle, S. Endo, A. Aspuru-Guzik, S. C. Benjamin, and X. Yuan, {\it Quantum computational chemistry}, 
Rev. Mod. Phys. {\bf 92}, 015003 (2020). 

 \bibitem{Bau20} Bela Bauer, Sergey Bravyi, Mario Motta, Garnet Kin-Lic Chan, {\it Quantum algorithms for quantum chemistry and quantum materials science}, Chem. Rev. {\bf 120}, 12685 (2020).


\bibitem{Bha22} K. Bharti et al., {\it Noisy intermediate-scale quantum (NISQ) algorithms}, Rev. Mod. Phys. {\bf 94}, 015004 (2022).  


%%%% general SB

\bibitem{Rin80} P. Ring and P. Schuck, {\it The Nuclear Many-Body Problem}
(Springer-Verlag, New-York, 1980).

\bibitem{Bla86} J. P. Blaizot and G. Ripka, {\it Quantum Theory of Finite Systems} (MIT Press, Cambridge, 1986).    


%%%% symmmetry restoration nuclear physics 


\bibitem{Ben03} M. Bender, P.-H. Heenen, and P.-G. Reinhard, {\it Self-consistent mean-field models for nuclear structure},  Rev. Mod. Phys. {\bf 75}, 121 (2003).

\bibitem{Rob18} L. M. Robledo, , T. R. Rodr\'iguez, and R. R. Rodr\'iguez-Guzm\'an, {\it Mean field and beyond description of nuclear structure with the Gogny force a review}, Journal of Physics G: Nuclear and Particle Physics {\bf 46},  013001 (2018).

\bibitem{She19}  J. A. Sheikh, J. Dobaczewski, P. Ring, L. M. Robledo, C. Yannouleas, {\it Symmetry restoration in mean-field approaches }, 
J. Phys. G: Nucl. Part. Phys {\bf 48}, 123001 (2021).

%%% Ab-initio symmetry breaking problem


%\bibitem{Kha21} Armin Khamoshi, Thomas Henderson, Gustavo Scuseria, {\it Correlating AGP on a quantum computer}, 
%Quantum Sci. Technol. {\bf 6}, 014004 (2021). 

%%% Symmetry protection for error corrections. 


\bibitem{Lac12} Denis Lacroix and Danilo Gambacurta, {\it  Projected quasiparticle perturbation theory}, Phys. Rev. C 86, 014306 (2012). 

\bibitem{Gam12} Danilo Gambacurta and Denis Lacroix, {\it Description of two-particle transfer in superfluid systems}
Phys. Rev. C 86, 064320 (2012)

\bibitem{Dug17} T Duguet and A Signoracci, {\it Symmetry broken and restored coupled-cluster theory: II. Global gauge symmetry and particle number}, J. Phys. G: Nucl. Part. Phys. 44, 015103 (2017). 

\bibitem{Qiu17} Yiheng Qiu, Thomas M. Henderson, Jinmo Zhao, and Gustavo E. Scuseria
, "Projected coupled cluster theory", J. Chem. Phys. 147, 064111 (2017). 

\bibitem{Qiu19} Y. Qiu, T. M. Henderson, T. Duguet, and G. E. Scuseria, {\it Particle-number projected Bogoliubov-coupled-cluster theory: Application to the pairing Hamiltonian},  Phys. Rev. C 99, 044301 (2019)


\bibitem{Rip17} J. Ripoche, D. Lacroix, D. Gambacurta, J.-P. Ebran, and T. Duguet, {\it Combining symmetry breaking and restoration with configuration interaction: A highly accurate many-body scheme applied to the pairing Hamiltonian}, Phys. Rev. C 95, 014326 (2017). 

\bibitem{Rip18} J. Ripoche, T. Duguet, J.-P. Ebran, and D. Lacroix, {\it Combining symmetry breaking and restoration with configuration interaction: Extension to z-signature symmetry in the case of the Lipkin model}, Phys. Rev. C 97, 064316 (2018). 


\bibitem{Fro21a} Mikael Frosini, Thomas Duguet, Jean-Paul Ebran, Vittorio Som\`a, {\it Multi-reference many-body perturbation theory for nuclei I -- Novel PGCM-PT formalism }, Eur. Phys. J. {\bf A 58}, 62 (2022). 

\bibitem{Fro21b} Mikael Frosini, Thomas Duguet, Jean-Paul Ebran, Benjamin Bally, Tobias Mongelli, Tom\' as R. Rodr\'iguez, Robert Roth, Vittorio Som\`a, {\it Multi-reference many-body perturbation theory for nuclei II -- Ab initio study of neon isotopes via PGCM and IM-NCSM calculations }, 
Eur. Phys. J. {\bf A 58}, 63 (2022). 

\bibitem{Fro21c} Mikael Frosini, Thomas Duguet, Jean-Paul Ebran, Benjamin Bally, Heiko Hergert, Tom\'as R. Rodríguez, Robert Roth, Jiangming Yao, Vittorio Som\`a, {\it Multi-reference many-body perturbation theory for nuclei III -- Ab initio calculations at second order in PGCM-PT}, Eur. Phys. J. {\bf A 58}, 64 (2022). 


\bibitem{Reg19} David Regnier and Denis Lacroix, {\it Microscopic description of pair transfer between two superfluid Fermi systems. II. Quantum mixing of time-dependent Hartree-Fock-Bogolyubov trajectories},
Phys. Rev. {\bf C 99}, 064615 (2019)

\bibitem{Ese93} C. Esebbag and J. L. Egido, {\it Number projected statistics and the pairing correlations at high excitation energies} Nucl. Phys. A 552, 205 (1993).

\bibitem{Gam13} Danilo Gambacurta, Denis Lacroix, and N. Sandulescu, {\it Pairing and specific heat in hot nuclei}, 
Phys. Rev. {\bf C 88}, 034324 (2013).
%%% SB in 

\bibitem{McA19} Sam McArdle, Xiao Yuan, and Simon Benjamin, {\it Error-Mitigated Digital Quantum Simulation}
Phys. Rev. Lett. 122, 180501 (2019).


\bibitem{Sch22}  Louis Schatzki, Martin Larocca, Frederic Sauvage, M. Cerezo, 
{\it Theoretical Guarantees for Permutation-Equivariant Quantum Neural Networks}, arXiv:2210.09974. 



%%%%%

\bibitem{Dir35} P. A. M. Dirac, {\it Quantum Mechanics} (Oxford University Press, 2nd ed. London 1935). 

\bibitem{Low70} Per‐Olov L\"owdin and Osvaldo Goscinski {\it The exchange phenomenon, the symmetric group, 
and the spin degeneracy problem}, {\it International Journal of Quantum Chemistry} vol. 4,  S3B, 533 (1969).

\bibitem{Hav19} Vojtech Havlicek, Sergii Strelchuk, and Kristan Temme, {\it Classical algorithm for quantum SU(2) Schur sampling}
Phys. Rev. A 99, 062336 (2019)

\bibitem{Siw21} P. Siwach and D. Lacroix,  {\it Filtering states with total spin on a quantum computer }, Phys. Rev. A 104, 062435 (2021) .  

\bibitem{Mar02} A. Marzuoli, M. Rasetti, {\it Spin network quantum simulator}, Physics Letters {\bf A 306} (2002).

\bibitem{Mar05} A. Marzuoli and M. Rasetti, {\it Computing spin networks}, Ann. Phys. (Amsterdam) {\bf 318}, 345 (2005).

\bibitem{Jor10} S. P. Jordan, {\it Permutational quantum computing}, Quantum Inf. Comput. {\bf 10}, 470 (2010).




%%% Jordan Wigner
\bibitem{Jor28} P. Jordan and E. Wigner, {\it \"Uber das Paulische \"Aquivalenzverbot},  Z. Phys. {\bf 47}, 631 (1928).

\bibitem{Lie61} Elliott Lieb, Theodore Schultz, Daniel Mattis, {\it Two soluble models of an antiferromagnetic chain}, Ann. of Phys. {\bf 16},  407 (1961).

 \bibitem{Som02} R. Somma, G. Ortiz, J. E. Gubernatis, E. Knill, and R. Laflamme, {\it Simulating physical phenomena by quantum networks},
Phys. Rev. {\bf A 65}, 042323 (2002).   

%%% Bravii Kitaek 

\bibitem{Bra02} S. Bravyi and A. Kitaev, {\it Fermionic quantum computation},  Ann. of Phys.  {\bf 298}, 210 (2002).

\bibitem{See12} J. T. Seeley, M. J. Richard, and P. J. Love, {\it The Bravyi-Kitaev transformation for quantum computation of electronic structure}, J. Chem. Phys. {\bf 137}, 224109 (2012).
 
%%% Simple discussion on symmetries 

\bibitem{Vat04} Farrokh Vatan and Colin Williams, {\it Optimal quantum circuits for general two-qubit gates}, 
Phys. Rev. {\bf A 69}, 032315 (2004). 

%%% Qiskit 

\bibitem{Abr19} {H{\'e}ctor Abraham \textit{et al} {[}Qiskit collaboration{]},
\textit{Qiskit: An Open-source Framework for Quantum Computing}, \textsf{https://qiskit.org/}
(2019), DOI: 10.5281/zenodo.2562110}


%%% Simple discussion on symmetries 

\bibitem{Gar20} B.T. Gard, L. Zhu, G.S. Barron, N. J. Mayhall, S. E. Economou and E. Barnes  {\it Efficient symmetry-preserving state preparation circuits for the variational quantum eigensolver algorithm}. npj Quantum Inf. {\bf 6}, 10 (2020). 

%%% circuits

\bibitem{Kay18} Alastair Kay, \textit{Tutorial on the Quantikz Package},
arXiv:1809.03842; DOI: 10.17637/rh.7000520

%%% Hamming weight 
\bibitem{HammingW} Glaisher, James Whitbread Lee, { \it On the residue of a binomial-theorem coefficient with respect to a prime modulus}. The Quarterly Journal of Pure and Applied Mathematics. {\bf 30}: 150–156, (1899).



\bibitem{Cer21} Michael J. Cervia, A. B. Balantekin, S. N. Coppersmith, Calvin W. Johnson, Peter J. Love, C. Poole, K. Robbins, and M. Saffman, {\it Lipkin model on a quantum computer}, Phys. Rev. {\bf C 104}, 024305 (2021). 

 
\bibitem{Kha21} A. Khamoshi, T. Henderson, and G. Scuseria, {\it Correlating AGP on a quantum computer}, Quant. Sci. Technol. {\bf 6}, 014004 (2021).

\bibitem{Rui22} Edgar Andres Ruiz Guzman and Denis Lacroix, {\it Accessing ground-state and excited-state energies in a many-body system after symmetry restoration using quantum computers}, Phys. Rev. {\bf C 105}, 024324 (2022).  

\bibitem{Aru20} F. Arute et al ( Google AI Quantum and Collaborators), {\it Hartree-Fock on a superconducting qubit quantum computer}, 
Science 369 (6507), 1084-1089, 2020



%%%  Thouless

\bibitem{Tho60} David J Thouless, {\it Stability conditions and nuclear rotations in the Hartree-Fock theory}, Nucl. Phys. {\bf 21}, 225 (1960).

%%% Unitary CC

\bibitem{Rom19} Jonathan Romero, Ryan Babbush, Jarrod R McClean, Cornelius Hempel, Peter J Love, and Al\'an Aspuru-Guzik, {\it Strategies for quantum computing molecular energies using the unitary coupled cluster ansatz}, Quantum Sci. Technol. {\bf 4}, 014008 (2019).

\bibitem{Guo21} Qing Guo and Ping-Xing Chen, {\it Optimization of VQE-UCC algorithm based on spin state symmetry}, Frontiers in Physics 9 (2021).


\bibitem{Ana22} Abhinav Anand, Philipp Schleich, Sumner Alperin-Lea, Phillip W. K. Jensen, Sukin Sim, Manuel D\'iaz-Tinoco, Jakob S. Kottmann, Matthias Degroote, Artur F. Izmaylov, Al\'an Aspuru-Guzik, {\it  A Quantum Computing View on Unitary Coupled Cluster Theory}, Chem. Soc. Rev. {\bf 51},1659 (2022). 


%%% Trotter-Suzuki 
\bibitem{Tro59} H. F. Trotter, Proc. Am. Math. Soc. 10, 545 (1959).

\bibitem{Kis22} Oriel Kiss, Michele Grossi, Pavel Lougovski, Federico Sanchez, Sofia Vallecorsa, Thomas Papenbrock, {\it  
Quantum computing of the $^6$Li nucleus via ordered unitary coupled cluster }, arXiv:2205.00864.  




%%% Nuclear

\bibitem{Dum18} E.F. Dumitrescu, A.J. McCaskey, G. Hagen, G. R. Jansen,
T.D. Morris, T. Papenbrock, R.C. Pooser, D.J. Dean, and P. Lougovski, {\it Cloud Quantum Computing of an Atomic Nucleus}, 
Phys. Rev. Lett. \textbf{120}, 210501 (2018).


%%%%%% Not cited %%%%%
%\bibitem{Lu19} Hsuan-Hao Lu, Natalie Klco, Joseph M. Lukens, Titus D. Morris, Aaina Bansal, Andreas Ekstr\"om, Gaute Hagen, Thomas Papenbrock, Andrew M. Weiner, Martin J. Savage, and Pavel Lougovski, {\it Simulations of subatomic many-body physics on a quantum frequency processor},
%Phys. Rev. A 100, 012320 (2019). 


\bibitem{Sek20} K. Seki, T. Shirakawa, and S. Yunoki, {\it Symmetry-adapted variational quantum eigensolver}, Phys. Rev. A 101, 052340 (2020).

\bibitem{Sek22}  Kazuhiro Seki, Seiji Yunoki, {\it Spatial, spin, and charge symmetry projections for a Fermi-Hubbard model on a quantum computer}, Phys. Rev. {\bf A 105}, 032419 (2022). 

%%% Clebsch Gordan and Schur transformation

\bibitem{Bac06} D. Bacon, I. L. Chuang, and A. W. Harrow, {\it Efficient Quantum Circuits for Schur and Clebsch-Gordan Transforms}, 
Phys. Rev. Lett. {\bf 97}, 170502 (2006).

\bibitem{Kir17} W. Kirby, {\it A practical quantum Schur transform, undergraduate thesis}, Williams College, 2017.

\bibitem{Kir18} W. M. Kirby and F. W. Strauch, {\it A practical quantum algorithm for the Schur transform}, Quantum Inf. Comput. {\bf 18}, 09 (2018).

\bibitem{Hav18} Vojtech Havlicek and Sergii Strelchuk, {\it Quantum Schur Sampling Circuits can be Strongly Simulated}, Phys. Rev. Lett. {\bf 121}, 060505 (2018).  

\bibitem{Kro19} Hari Krovi, {\it An efficient high dimensional quantum Schur transform}, Quantum {\bf 3}, 122 (2019).  


%%%% Spin projection

\bibitem{Sug16} Kenji Sugisaki, Satoru Yamamoto, Shigeaki Nakazawa, Kazuo Toyota, Kazunobu Sato, Daisuke Shiomi, and Takeji Takui, 
{\it Quantum Chemistry on Quantum Computers: A Polynomial-Time Quantum Algorithm for Constructing the Wave Functions of Open- Shell Molecules}, 
J. of Phys. Chem. {\bf A 120} (32), 6459 (2016). 

\bibitem{Sug19} K. Sugisaki, S. Yamamoto, S. Nakazawa, K. Toyota, K. Sato, D. Shiomi, and T. Takui, {\it Open shell electronic state calculations on quantum computers: A quantum circuit for the preparation of configuration state functions based on Serber construction }, Chem. Phys. Lett. {\bf 737S}, 
100002 (2019).


\bibitem{Lui19} J.-G. Liu, Y.-H. Zhang, Y. Wan, and L. Wang, {\it Variational quantum eigensolver with fewer qubits}, Phys. Rev. Res. {\bf 1}, 023025 (2019).

%\bibitem{Sek20}  Kazuhiro Seki, Tomonori Shirakawa, and Seiji Yunoki, {\it Symmetry-adapted variational quantum eigensolver}
%Phys. Rev. {\bf A 101}, 052340 (2020).

\bibitem{Hla22} Manqoba Q. Hlatshwayo, Yinu Zhang, Herlik Wibowo, Ryan LaRose, Denis Lacroix, Elena Litvinova, 
{\it Simulating excited states of the Lipkin model on a quantum computer}, arXiv:2203.01478 

\bibitem{Lip65} H. J. Lipkin, N. Meshkov, and A. Glick, {\it Validity of many-body approximation methods for a solvable model:(i) Exact solutions and perturbation theory}, Nucl. Phys. {\bf B 62}, 188 (1965).

%\bibitem{Cer21} Michael J. Cervia, A. B. Balantekin, S. N. Coppersmith, Calvin W. Johnson, Peter J. Love, C. Poole, K. Robbins, and M. Saffman, 
%{\it Lipkin model on a quantum computer}, Phys. Rev. {\bf C 104}, 024305 (2021). 


%%% Pairing Ansatz


\bibitem{Bri05}  D. M. Brink and R. A. Broglia, {\it Nuclear Superfluidity: Pairing in Finite Systems} (Cambridge University Press, 2005).


\bibitem{Ver09} F. Verstraete, J. I. Cirac, and J. I. Latorre,  {\it Quantum circuits for strongly correlated quantum systems} Phys. Rev. {\bf A 79}, 032316 (2009).

 \bibitem{Jia18} Zhang Jiang, Kevin J. Sung, Kostyantyn Kechedzhi, Vadim N. Smelyanskiy, and Sergio Boixo, {\it  Quantum Algorithms to Simulate Many-Body Physics of Correlated Fermions}, Phys. Rev. Applied {\bf 9}, 044036 (2018).


\bibitem{Lac20} Denis Lacroix, {\it Symmetry-Assisted Preparation of Entangled Many-Body States on a Quantum Computer}, Phys. Rev. Lett. {\bf 125}, 230502 (2020). 


\bibitem{Kha22} Armin Khamoshi, Guo P. Chen, Francesco A. Evangelista, Gustavo E. Scuseria, {\it AGP-based unitary coupled cluster theory for quantum computers}, arxiv:2205.13420

\bibitem {Rui21} E.A. Ruiz Guzman and D. Lacroix,{\it Calculation of generating function in many-body systems with quantum computers: technical challenges and use in hybrid quantum-classical methods}, arXiv:2104.08181. 


%%%%

 \bibitem{Saw16} N. P. D. Sawaya, M. Smelyanskiy, J. R. McClean, and A. Aspuru-Guzik, {\it Error Sensitivity to Environmental Noise in Quantum Circuits for Chemical State Preparation} J. Chem. Theory Comput. 12, 3097 (2016).
 
  
%\bibitem{McA19} Sam McArdle, Xiao Yuan, and Simon Benjamin, {\it Error-Mitigated Digital Quantum Simulation}
%Phys. Rev. Lett. 122, 180501 (2019).


\bibitem{Got97} D. Gottesman, {\it Stabilizer Codes and Quantum Error Correction}, Caltech Ph.D. Thesis, arXiv:quant-ph/9705052.

 
\bibitem{Bon18} X. Bonet-Monroig, R. Sagastizabal, M. Singh, and T. E. O'Brien, {\it Low-cost error mitigation by symmetry verification}, 
Phys. Rev. A 98, 062339 (2018) 

\bibitem{Sag19} R. Sagastizabal, X. Bonet-Monroig, M. Singh, M. A. Rol, C. C. Bultink, X. Fu, C. H. Price, V. P. Ostroukh, N. Muthusubramanian, A. Bruno, M. Beekman, N. Haider, T. E. O'Brien, and L. DiCarlo, {\it Experimental error mitigation via symmetry verification in a variational quantum eigensolver}, Phys. Rev. A 100, 010302(R) (2019) 


\bibitem{Tra21} Minh C. Tran, Yuan Su, Daniel Carney, and Jacob M. Taylor, {\it Faster Digital Quantum Simulation by Symmetry Protection}
PRX Quantum 2, 010323 (2021). 

\bibitem{Koc21} B. Koczor, Exponential Error Suppression for Near-Term Quantum Devices, Phys. Rev. {\bf X 11}, 031057 (2021). 

\bibitem{Hug21} William J. Huggins, Sam McArdle, Thomas E. O’Brien, Joonho Lee, Nicholas C. Rubin, Sergio Boixo, K. Birgitta Whaley, Ryan Babbush, and Jarrod R. McClean, {\it Virtual Distillation for Quantum Error Mitigation},  Phys. Rev. {\bf X 11}, 041036 (2021).  



%%% 
\bibitem{Yen19} Tzu-Ching Yen, Robert A. Lang, Artur F. Izmaylov,  {\it Exact and approximate symmetry projectors for the electronic structure problem on a quantum computer}, Chem. Phys. 151, 164111 (2019)

\bibitem{Izm19}  A. F. Izmaylov, {\it On construction of projection operators},  J. Phys. Chem. A 123, 3429 (2019).

\bibitem{Low55} Per-Olov L\"owdin, {\it Quantum Theory of Many-Particle Systems. III. Extension of the Hartree-Fock Scheme to Include 
Degenerate Systems and Correlation Effects},  Phys. Rev. 97, 1509 (1955). 

\bibitem{Low69} Per‐Olov L\"owdin and Osvaldo Goscinski {\it The exchange phenomenon, the symmetric group, 
and the spin degeneracy problem}, {\it International Journal of Quantum Chemistry} vol. 4,  S3B, 533 (1969).

\bibitem{Mol16} N. Moll, A. Fuhrer, P. Staar, and I. Tavernelli, {\it Optimizing qubit resources for quantum chemistry simulations in second quantization on a quantum computer}, J. Phys. A: Math. Theor. {\bf 49}, 295301 (2016).


%%% Discretization 

\bibitem{Fom70} V. N. Fomenko, {\it Projection in the occupation-number space and the canonical transformation}, J. Phys. G 3, 8 (1970).

\bibitem{Ben09} M. Bender, T. Duguet, and D. Lacroix, {\it Particle-number restoration within the energy density functional formalism}, 
Phys. Rev. C 79, 044319 (2009). 

%%%% LCU method 

%% Original idea LCU

\bibitem{Lon06} ] G.L. Long, {\it General quantum interference principle and duality computer}, Common. Theor. Phys. {\bf 45}, 825-844, (2006).

%% With B^\dagger

\bibitem{Chi12}  A. M. Childs and N. Wiebe, {\it Hamiltonian simulation using linear combinations of unitary operations}, 
Quantum Inf. Comput. 12, 901 (2012).

\bibitem{Ber14} Dominic W. Berry, Andrew M. Childs, Richard Cleve, Robin Kothari, and Rolando D. Somma, {\it 
Exponential improvement in precision for simulating sparse Hamiltonians}, Proceedings of the 46th ACM Symposium on Theory of Computing (STOC 2014), 283 (2014). arXiv:1312.1414

\bibitem{Ber15} Dominic W. Berry, Andrew M. Childs, and Robin Kothari, {\it Hamiltonian Simulation with Nearly Optimal Dependence on all Parameters},
Proceedings of the 56th IEEE Symposium on Foundations of Computer Science (FOCS 2015), 792 (2015). arXiv:1501.01715

%% With Hadamard

\bibitem{Wei20} S. Wei, H. Li, and G. Long, {\it A Full Quantum Eigensolver for Quantum Chemistry Simulations}. Research, {\bf 2020},(2020).

%%% Spin as LCU 

\bibitem{Tsu20}  T. Tsuchimochi, Y. Mori, and S. L. Ten-no, {\it Spin-projection for quantum computation: A low-depth approach to strong correlation}, Phys. Rev. Research 2, 043142 (2020).

\bibitem{Tsu22} Takashi Tsuchimochi, Masaki Taii, Taisei Nishimaki, Seiichiro L. Ten-no, {\it Adaptive construction of shallower quantum circuits with quantum spin projection for fermionic systems}, arXiv:2205.07097.  


%%% Old refs on spins

\bibitem{Ata73}, B. Atalay, D.M. Brink and A. Mann, {\it A product form for projection operators}, Phys. Lett. {\bf 46B}, 145 (1973). 

\bibitem{Sme73} Y. G. Smeyers and L. Doreste-Suarez, {\it Half-Projected and Projected Hartree-Fock Calculations for Singlet Ground States. I. Four-Electron Atomic Systems},  Int. J. Quantum Chem. {\bf 7}, 687 (1973).


%%% Oracle and Grover 

\bibitem{Gro97a} Lov K. Grover, {\it Quantum Mechanics Helps in Searching for a Needle in a Haystack},  Phys. Rev. Lett. {\bf 79}, 325 (1997). 

\bibitem{Gro97b} Lov K. Grover, {\it Quantum Computers Can Search Arbitrarily Large Databases by a Single Query}, Phys. Rev. Lett. {\bf 79}, 4709 (1997) 


\bibitem{Kay11}  Phillip Kaye, Raymond Laflamme and Michele Mosca,{\it An Introduction to Quantum Computing},  Oxford University Press,  Oxford (2011).

\bibitem{Lim19} Franklin de Lima Marquezino, Renato Portugal, et al., {\it A Primer on Quantum Computing},  Springer Nature Switzerland (2019).

%%% Trotter Suzuki

%\bibitem{Tro59} H. F. Trotter, {\it On the product of semi-groups of operators} Proc. Am. Math. Soc. {\bf 10}, 545 (1959).

 
\bibitem{Bab15}  R. Babbush, J. McClean, D. Wecker, A. Aspuru-Guzik, and N. Wiebe,{\it  Chemical basis of Trotter-Suzuki errors in quantum chemistry simulation}, Phys. Rev. {\bf A 91}, 022311 (2015).


%%%% General 

\bibitem{Nie02} M. A. Nielsen and I. L. Chuang. {\it Quantum information and quantum computation.}, Cambridge University Press (2000). 
 
%%% Some numerical method
\bibitem{She06} Vivek V. Shende, Stephen S. Bullock, Igor L. Markov, {\it Synthesis of Quantum Logic Circuits}, 
IEEE Trans. on Computer-Aided Design, {\bf 25}, 1000, (2006).

%%%LCU
\bibitem{Siw02} Pooja Siwach, P. Arumugam {\it Quantum computation of nuclear observables involving linear combination of unitary operators}, Phys. Rev. C {\bf 105}, 064318 (2022).


%%%% Walsh decomposition


\bibitem{Wal23} J. L. Walsh, {\it A closed set of normal orthogonal functions}, Am. J. Math. {\bf 45}, 5 (1923). 
 
 \bibitem{Wel14} Jonathan Welch, Daniel Greenbaum, Sarah Mostame, and Alan Aspuru-Guzik, {\it Efficient quantum circuits for diagonal unitaries without ancillas},  New J. Phys. {\bf 16}, 033040 (2014) 
 
\bibitem{Hoy00} Peter Hoyer, {\it On Arbitrary phases in quantum amplitude amplification}, Phys. Rev. {\bf A 62}, 052304 (2000)
 
%\bibitem{Sch03}
%Norbert Schuch and Jens Siewert, {\it Programmable Networks for Quantum Algorithms}, Phys. Rev. Lett. {\bf 91}, 027902 (2003).

%\bibitem{Bul03} S.S. Bullock and I. L. Markov, {\it Asymptotically optimal circuits for arbitrary n-qubit diagonal computations}, 
%Quantum Inf. Comput. {\bf 4},  27 (2004). 


%%%REF TO CHECK NO DOUBLE


\bibitem{Hid19} J. D. Hidary, {\it Quantum Computing: An Applied Approach}, Springer International Publishing, (2019).


%\bibitem{Fan19} Guido Fano, S. M. Blinder, Mathematical Physics in Theoretical Chemistry, 377 (2019).

%%%%% No cited 
%\bibitem{Ovr03} E. Ovrum. {\it Quantum computing and many-body physics}. Master's thesis, University of Oslo, (2003).
%\bibitem{Ovr07} Ovrum E, Hjorth-Jensen M. {\it Quantum computation algorithm for many-body studies}, arXiv:0705.1928v1.
%\bibitem{Per14} Alberto Peruzzo, Jarrod McClean, Peter Shadbolt, Man-Hong Yung, Xiao-Qi Zhou, Peter J. Love, Al\'an Aspuru-Guzik and Jeremy L. O'Brien, {\it A variational eigenvalue solver on a photonic quantum processor}, Nat. Commun. {\bf 5}, 4213 (2014). 
%\bibitem{McC16} Jarrod R McClean, Jonathan Romero, Ryan Babbush and Al\'an Aspuru-Guzik, {\it The theory of variational hybrid %quantum-classical algorithms},  New J. of Phys. {\bf 18},  023023 (2016). 

%%% Kitaev


\bibitem{Kit95}  A. Y. Kitaev, {\it Quantum measurements and the Abelian Stabilizer Problem} arXiv:quant-ph/9511026.

%% IQPE
\bibitem{Dob07} M. Dob\ifmmode \check{s}\else \v{s}\fi{}\'{\i}\ifmmode \check{c}\else \v{c}\fi{}ek, G. Johansson, V. Shumeiko, and G. Wendin, {\it Arbitrary accuracy iterative quantum phase estimation algorithm using a single ancillary qubit: A two-qubit benchmark}, Phys. Rev. A {\bf 76}, (2007).



\bibitem{Lab21} M. L. LaBorde, Mark M. Wilde, {\it Testing symmetry on quantum computers}, arXiv:2105.12758

\bibitem{Lab22}  M. L. LaBorde, Mark M. Wilde, {\it  Quantum Algorithms for Testing Hamiltonian Symmetry}, arXiv:2203.10017

\bibitem{Sil22} A. J. da Silva and Daniel K. Park, {\it Linear-depth quantum circuits for multiqubit controlled gates}, Phys. Rev. A {\bf 106}, (2022).


%%% Quantum State Preparation

\bibitem{Moz22} F. Mozafari, G. De Micheli, and Y. Yang, {\it Efficient deterministic preparation of quantum states using decision diagrams} Phys. Rev. A {\bf 106}, (2022).

\bibitem{She04} V. V. Shende, I. L. Markov, and S. S. Bullock, {\it Minimal universal two-qubit controlled-NOT-based circuits}, Phys. Rev. A {\bf 69}, (2004).

\bibitem{Min21} X. Zhang, M. Yung, and X. Yuan, {\it Low-depth quantum state preparation}, Phys. Rev. Research {\bf 3}, (2021).

\bibitem{Ara21} I.F. Araujo, D.K. Park, F. Petruccione, {\it et al}, {\it A divide-and-conquer algorithm for quantum state preparation}. Sci Rep {\bf 11}, (2021). 

%%% Diagonal Operators

\bibitem{Bul04} S.S. Bullock and I.L. Markov, {\it Smaller Circuits for Arbitrary n-qubit Diagonal Computations}, Quant. Inf. and Comp., {\bf 4}, (2004).

%% Quantum Fourier Transform

%\bibitem{Hal00} L. Hales and S. Hallgren, {\it An improved quantum Fourier transform algorithm and applications}, Proceedings 41st Annual Symposium on Foundations of Computer Science, (2000).

%\bibitem{Nam20} Y. Nam, Y. Su, and D. Maslov, {\it Approximate quantum Fourier transform with $\mathcal{O}\left(n log\left(n\right)\right)$ T gates}. npj Quantum Inf {\bf 6}, (2020).


\end{thebibliography}
\end{document}